\documentclass[journal,twoside]{IEEEtran}

\usepackage{graphicx}
\usepackage{epsfig} 
\usepackage{times} 
\usepackage{amsmath} 
\usepackage{amssymb}  
\usepackage{url}
\usepackage{algorithm,algorithmic}
\usepackage{cases}
\usepackage{cite}
\usepackage{enumerate}

\usepackage{dsfont}
\usepackage{slashbox}

\ifCLASSOPTIONcompsoc
\usepackage[caption=false,font=normalsize,labelfont=sf,textfont=sf]{subfig}
\else
\usepackage[caption=false,font=footnotesize]{subfig}

\usepackage{theorem}
\theorembodyfont{\rmfamily}

\newtheorem{prob}{Problem}
\newtheorem{lem}{Lemma}
\newtheorem{defin}{Definition}
\newtheorem{theorem}{Theorem}
\newtheorem{prop}{Proposition}
\newtheorem{assum}{Assumption}

\renewcommand{\IEEEQED}{\IEEEQEDopen}

\newcommand{\ol}[1]{\overline{#1}}
\newcommand{\hs}{& \hspace{-3mm}}

\newcommand{\tl}[1]{\tilde{#1}}

\begin{document}
\title{Disconnection-aware Attack Detection and Isolation with Separation-based Detector Reconfiguration}

\author{Hampei~Sasahara,~\IEEEmembership{Member,~IEEE,}
        Takayuki~Ishizaki,~\IEEEmembership{Member,~IEEE,}
        Jun-ichi~Imura,~\IEEEmembership{Senior~Member,~IEEE,}
        and~Henrik~Sandberg,~\IEEEmembership{Senior~Member,~IEEE}
\thanks{Manuscript received Xxx xx, 20xx; revised Xxx xx, 20xx. This work was funded in part by the Swedish Research Council (project 2016-00861) and KTH Digital Futures (project DEMOCRITUS).}
\thanks{H. Sasahara and H. Sandberg are with the Division of Decision and Control Systems, KTH Royal Institute of Technology, Stockholm, SE-100 44 Sweden email:\{hampei, hsan\}@kth.se.}
\thanks{T. Ishizaki and J. Imura are with the Graduate School of Engineering, Tokyo Institute of Technology, Tokyo, 152-8552 Japan e-mail:\{ishizaki, imura\}@sc.e.titech.ac.jp.}
\thanks{\copyright 2021 IEEE.  Personal use of this material is permitted.  Permission from IEEE must be obtained for all other uses, in any current or future media, including reprinting/republishing this material for advertising or promotional purposes, creating new collective works, for resale or redistribution to servers or lists, or reuse of any copyrighted component of this work in other works.}
}

\markboth{IEEE TCST,~Vol.~xx, No.~xx, Xxx~2021}%
{Sasahara \MakeLowercase{\textit{et al.}}: Disconnection-aware Attack Detection and Isolation with Separation-based Detector Reconfiguration}

\maketitle

\begin{abstract}
This study addresses incident handling during an adverse event for dynamical networked control systems.
Incident handling can be divided into five steps: detection, analysis, containment, eradication, and recovery.
For networked control systems, the containment step can be conducted through physical disconnection of an attacked subsystem.
In accordance with the disconnection, the equipped attack detection unit should be reconfigured to maintain its detection capability.
In particular, separating the detection subunit associated with the disconnected subsystem is considered as a specific reconfiguration scheme in this study.
This paper poses the problem of disconnection-aware attack detection and isolation with the separation-based detector reconfiguration.
The objective is to design an attack detection unit that preserves its detection and isolation capability even under any possible disconnection and separation.
The difficulty arises from network topology variation caused by disconnection that can possibly lead to stability loss of the distributed observer inside the attack detection unit.
A solution is proposed based on an existing controller design technique referred to as retrofit control.
Furthermore, an application to low-voltage power distribution networks with distributed generation is exhibited.
Numerical examples evidence the practical use of the proposed method through a benchmark distribution network.
\end{abstract}


%
\IEEEpeerreviewmaketitle

\section{Introduction}

\IEEEPARstart{T}{he} recent tremendous drive towards increasing connectivity among cyber-physical components leaves the resulting networked systems vulnerable to adversarial attacks.
In fact, substantive malware programs targeting physical systems have been reported~\cite{History2018Hemsley},
and some of them, such as, Stuxnet~\cite{Nicolas2011Stuxnet,CISA2014}, BlackEnergy~3~\cite{CISA2018}, and HatMan~\cite{CISA2017}, have succeeded in causing serious damages to critical infrastructure networks~\cite{David2015Security}.
For secure operation of networked physical systems, novel security schemes in the physical layer are required in addition to the existing information security techniques.
This is mainly because of the difference between the requirements of information systems and physical systems.
For instance, real-time constraints, complexity, feedback, and legacy devices with limited computational power are major obstacles for physical systems~\cite{Alvaro2011Attacks}.
Furthermore, enhancing security in physical layers as well as information layers fits the notion of ``defense in depth'' advocated in~\cite{Kulpers2006Control}, which argues the importance of duplex protections.
For an overview of control system security, see~\cite{Seyed2019Systems}.

Model-based attack detection~\cite{Giraldo2018A} is one of the most used techniques in the protection schemes provided by the control community.
The basic idea is to create a dynamical model that imitates the evolution of physical state and to confirm that data collected from the actual system coincide with the predicted time series.
Typically, an attack detection unit is composed of a residual generator and an attack detector.
The residual generator calculates the discrepancy between the measured output and the predicted output,
while the attack detector decides whether to raise an alarm based on the residual signal exploiting its statistics.
The detection unit can possess an additional function of isolation, namely, identifying the components being attacked~\cite{Pasqualetti2013Attack}.
Classical model-based fault diagnosis techniques~\cite{Ding2013Model} help in residual generator design,
and hypothesis testing methods can be utilized for designing an attack detector~\cite{Murguia2019On}.

Meanwhile, according to the security guide for information systems provided by National Institute of Standards and Technology~\cite{Paul2012Computer},
\emph{incident handling} during an adverse event can be divided into five steps: detection, analysis, containment, eradication, and recovery.
In particular, containment is conducted by disconnecting a segment of infected workstations from the network~\cite{Paul2012Computer,Patrick2012Incident}.
When this idea is analogized to networked control systems, the model-based attack detection can be regarded as the detection and analysis steps,
and the containment step can be performed through physical disconnection of infected subsystems from the entire network.
During the incident handling, the equipped attack detection unit should be reconfigured in accordance with network topology variation caused by disconnection for containment.
As a specific reconfiguration scheme, we adopt \emph{separation-based reconfiguration}, namely,
separating the local detection subunit associated with the disconnected subsystem without modification of the remaining units.
This reconfiguration can be quickly conducted by simply switching off the corresponding communication.
Further, this scheme can easily be implemented because neither a bank of pre-designed detection units nor redesign of those units are required.


This study addresses the disconnection-aware attack detection and isolation problem with separation-based detector reconfiguration for incident handling in networked control systems.
The technical difficulty arises from its variable communication topology, which drastically changes the dynamics of the entire residual generator.
Typically, the residual generator contains a state observer with output estimation error feedback on the premise that the system to be protected is fixed.
This feedback architecture of the residual generator can possibly violate its stability under separation-based reconfiguration.
Thus, its detection capability cannot be guaranteed even if the residual generator can operate well for the nominal system where all subsystems are connected.

This paper proposes a residual generator design method that can preserve its stability and tracking capability under any disconnection.
The proposed approach is to preserve the entire stability by designing a distributed observer for every local component.
This idea is borrowed from an existing controller design technique referred to as retrofit control~\cite{Ishizaki2018Retrofit,Ishizaki2019Modularity,Sasahara2021Parameterization}, which has originally been proposed for modular design of control systems.
It is shown that, the proposed architecture of the residual generator can preserve not only the stability but also the detection capability.
It is also shown that attack isolation can also be performed through the proposed architecture under disconnection.
Furthermore, an application to inverter-based low-voltage distribution networks with distributed generation is exhibited.



\subsection*{Contribution}

First, we formally pose the disconnection-aware attack detector design problem in the context of incident handling.
Second, we provide a design method of a detection unit that can preserve its detection capability even under separation-based reconfiguration.
Further, we propose a disconnection-aware isolation filter design method.
Third, we demonstrate its application to low-voltage distribution networks with distributed energy resources.
Fourth, and finally, we illustrate the potential impact of our theoretical development through compelling examples.
In particular, we numerically confirm the effectiveness of the designed attack detector using a benchmark model of a European distribution network~\cite{Strunz2014Benchmark}.
A preliminary version of this work was presented in~\cite{Sasahara2020Disconnection}.
The additional topics include design of isolation filters, detailed proofs of the theoretical findings, and elaborate simulation results.

\subsection*{Related Work}

A few related works that propose secure control system design with separation-based reconfiguration can be found.
To the best of the authors' knowledge, fallback control in~\cite{Sasaki2015Model,Sasaki2017Model} is the first work that explicitly points out the importance of continuous operation of control systems under attack containment.
The fallback system is designed so as to enable the protected system to operate without communication from the external network,
and then fundamental functions are not lost even under containment of attacks.
A possible drawback is that the design method is inapplicable to complex systems because a switched Lyapunov function, which could be difficult to find for large-scale systems, is needed to be designed.
Another potentially applicable approach is the plug-and-play distributed fault detection~\cite{Boem2019Distributed} based on the partition-based distributed Kalman filter~\cite{Farina2018Partition}.
The residual generator design can be carried out in a distributed and scalable manner as long as the interaction matrices satisfy a small-gain condition.
Similarly, based on the small-gain approach, a distributed residual generator which can be designed only with its corresponding local subsystem is proposed in~\cite{Pasqualetti2013Attack}.
A potential shortcoming of those approaches is inapplicability to strongly interconnected systems.

Active fault tolerant control~\cite{Blanke2016Diagnosis} is a promising approach to handle drastic change of the system dynamics to be controlled, including a residual generator in the context of diagnosis.
In its framework, the dynamics of the designed controller is adjusted in response to component malfunctions.
Reconfiguration without physical redundancy can be classified into the following threefold~\cite{Lunze2008Reconfigurable}: projection with a bank of pre-designed controllers, learning, and automatic redesign.
In any case, large memory or powerful processing units are required for implementation, which can be restrictive in highly complex systems.
Moreover, the reconfiguration should be quickly carried out especially in the presence of a strategic attacker.
Separation-based reconfiguration in our approach does not need such abundant computational resources and can be conducted immediately after attack detection.

\subsection*{Organization and Notation}

In Section~\ref{sec:prob}, we provide a mathematical model of the networked system to be protected and the attack detection unit to be designed.
Based on the preliminaries, the disconnection-aware attack detection and isolation problems are formulated.
Section~\ref{sec:sol} solves the formulated problems.
In Section~\ref{sec:app}, we demonstrate the proposed design procedure for a low-voltage distribution network with distributed generation.
Section~\ref{sec:sim} verifies the theoretical findings and the practical effectiveness of our proposed approach through numerical examples for the CIGRE (International Council on Large Electric Systems) benchmark model~\cite{Strunz2014Benchmark}.
Finally, Section~\ref{sec:conc} draws conclusion.

The cardinality of a set $\mathcal{I}$ is denoted by $|\mathcal{I}|$,
the power set of a set $\mathcal{X}$ is denoted by $2^{\mathcal{X}}$,
the dimension of a vector $x$ by ${\rm dim}(x)$,
the transpose of a matrix $M$ by $M^{\sf T}$,
the rank of a matrix $M$ by ${\rm rank}\,M$,
the vector where $x_i$ for $i\in \mathcal{I}$ are concatenated vertically by $x_{\mathcal{I}}$,
the block diagonal system whose diagonal blocks are composed of $G_i$ for $i\in\mathcal{I}$ by ${\rm diag}(G_i)_{i\in\mathcal{I}}$,
where the subscript is omitted when $\mathcal{I}$ is clear from the context,
the set of all real rational transfer function matrices by $\mathcal{R}$,
the set of all stable real rational transfer function matrices by $\mathcal{RH}_{\infty}$,
and the normal rank of a transfer matrix $G$~\cite{Kailath1980Linear} by ${\rm rank}\,G$.

\section{Problem Formulation: Disconnection-aware Attack Detection and Isolation}
\label{sec:prob}

\subsection{Networked Control System, Model-based Attack Detector, and Separation-based Detector Reconfiguration}

Consider a networked control system being possibly under attack.
Let a dynamical model of the networked system be given as a linear time-invariant system composed of $N$ subsystems
\begin{equation}\label{eq:ori_sys}
 \Sigma_i:\left\{
 \begin{array}{cl}
 \dot{x}_i \hs = A_i x_i + B_i r_i + U_i v_i + X_ia_i\\
 y_i \hs = C_i x_i +D_i r_i + V_i v_i+Y_ia_i\\
 w_i \hs = E_ix_i + F_i r_i + W_i v_i+Z_ia_i
 \end{array}
 \right.
 ,\quad i=1,\ldots,N
\end{equation}
where $x_i,r_i,y_i,v_i,w_i,a_i$ denote the state, the reference input, the measurement output, the inflowing interaction, the outflowing interaction, and the signal caused by the attack, respectively.
Let
\[
 \left[
 \begin{array}{c}
 y_i\\
 w_i
 \end{array}
 \right]=
 \left[
 \begin{array}{lll}
 G_{y_i r_i} & G_{y_i v_i} & G_{y_i a_i}\\
 G_{w_i r_i} & G_{w_i v_i} & G_{w_i a_i}\\
 \end{array}
 \right]
 \left[
 \begin{array}{c}
 r_i\\
 v_i\\
 a_i
 \end{array}
 \right]
\]
denote the frequency-domain representation of the $i$th subsystem.
The interaction among the subsystems is represented by
\begin{equation}\label{eq:ori_interaction_nominal}
v = Lw
\end{equation}
with a transfer matrix $L$ where $v$ and $w$ are the stacked vectors of $v_i$ and $w_i$ for $i=1,\ldots,N$, respectively.
The networked control system is assumed to be well-posed~\cite[Definition~5.1]{Zhou1996Robust}.

\begin{figure}[t]
\centering
\includegraphics[width=.98\linewidth]{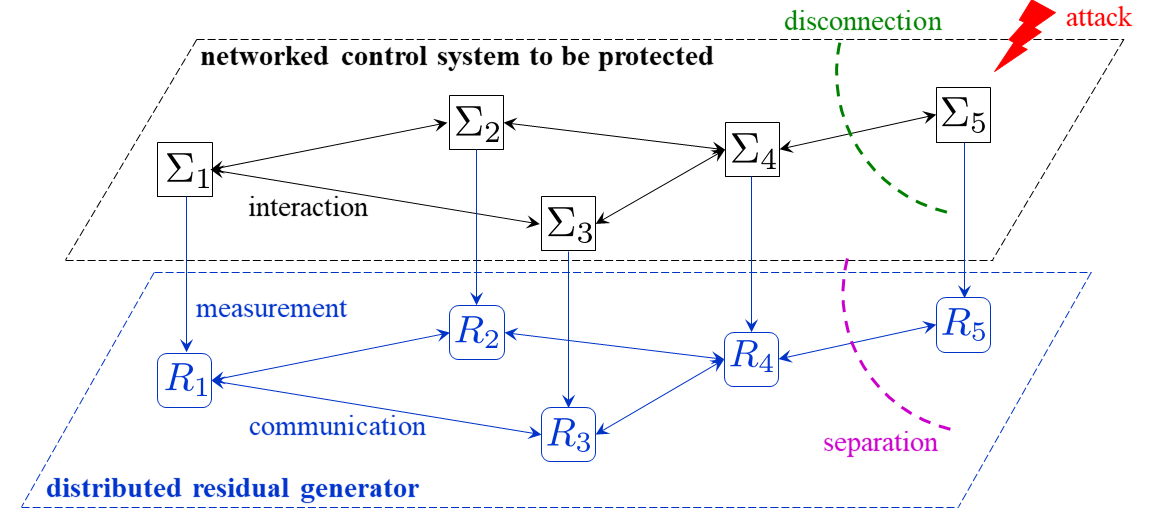}
\caption{Architecture of the networked control system to be protected and the distributed residual generator to be designed.
The distributed residual generator has the same network topology as that of the networked control system.
In this figure, when an alarm rings at the fifth detection unit, the corresponding subsystem $\Sigma_5$ is disconnected from the networked system for attack containment.
Simultaneously, the distributed residual generator is reconfigured through the separation of the fifth local residual generator $R_5$.
}
\label{fig:sys_structure}
\end{figure}

We consider designing a model-based attack detector with measurement of the outputs and information on the reference signals using the dynamical model.
A typical architecture of an attack detection unit is composed of a residual generator, which calculates the discrepancy between the measured output and the predicted output, and an attack detector, which decides whether to raise an alarm based on the residual signal.
For scalable implementation, we impose a distributed structure on the residual generator to be designed such that
\begin{equation}\label{eq:local_residual_generator}
 R_i:
 \left[
 \begin{array}{c}
 \epsilon_i\\
 \hat{w}_i
 \end{array}
 \right]=
 \left[
 \begin{array}{lll}
 R_{\epsilon_i y_i} & R_{\epsilon_i r_i} & R_{\epsilon_i \hat{v}_i}\\
 R_{\hat{w}_i y_i} & R_{\hat{w}_i r_i} & R_{\hat{w}_i \hat{v}_i}
 \end{array}
 \right]
  \left[
 \begin{array}{c}
 y_i\\
 r_i\\
 \hat{v}_i
 \end{array}
 \right]
\end{equation}
for $i=1,\ldots,N$, where $\epsilon_i$ denotes the $i$th residual signal, with the communication system
\begin{equation}\label{eq:communication_nominal}
 \hat{v}=\hat{L}\hat{w}
\end{equation}
where $\hat{v}$ and $\hat{w}$ are the stacked vectors of communication signals $\hat{v}_i$ and $\hat{w}_i$ transmitted through a transfer matrix $\hat{L}$.
The transfer matrix $\hat{L}$ is assumed to have the same sparsity pattern as that of $L$, i.e., the $(i,j)$th block component of $\hat{L}$ is zero if that of $L$ is zero.
Its architecture is illustrated by Fig.~\ref{fig:sys_structure}, where the distributed residual generator has the same network topology as that of the networked control system.
Based on the generated residual signal, an attack detector decides whether to raise an alarm, i.e.,
\begin{equation}\label{eq:detector}
 \theta_i = \Theta_i(\epsilon_i),\quad i=1,\ldots,N
\end{equation}
where $\theta_i(t)$, which takes a binary value, represents the decision at time $t$ and $\Theta_i$ denotes the decision rule.
The decision rule can be either static or dynamic,
where static detectors are often referred to as \emph{stateless detectors,} while dynamic detectors are referred to as \emph{stateful detectors.}

The above processes correspond to the detection and analysis steps of incident handling~\cite{Paul2012Computer}.
The next step is containment of attacks for reducing their impacts before the effects spread over the network.
In practical cyber-physical systems, there are several options for containment, including disconnection from a network, replacement with a redundant device, shutting down a workstation, and disabling certain functions~\cite{Young2016Incident}.
In this study, disconnection of components presumed to be attacked is treated as a specific action for containment.
We suppose that, when the $i$th attack detector raises an alarm, a collection of subsystems including the $i$th subsystem $\Sigma_i$ is disconnected from the networked control system.


Network topology change caused by disconnection leads to variation of the dynamics.
Let $\mathcal{I}\subset \{1,\ldots,N\}$ denote the index set of the remaining subsystems after the disconnection.
The interaction among the remaining subsystems is assumed to be represented by
\begin{equation}\label{eq:ori_interaction}
v_{\mathcal{I}} = L_{\mathcal{I}}w_{\mathcal{I}}
\end{equation}
where $L_{\mathcal{I}}$ denotes the submatrix of $L$ composed of its $(i,j)$th block components for $i,j\in \mathcal{I}$.
The input-output map can be represented by
\begin{equation}\label{eq:map_ra_to_y}
 y_{\mathcal{I}} = T_{y_\mathcal{I} r_\mathcal{I}}r_\mathcal{I} + T_{y_\mathcal{I} a_\mathcal{I}}a_\mathcal{I}
\end{equation}
where
\[
 \begin{array}{l}
 T_{y_\mathcal{I} r_\mathcal{I}}:={\rm diag}(G_{y_ir_i})+{\rm diag}(G_{y_iv_i})Q_{\mathcal{I}}{\rm diag}(G_{w_ir_i}),\\
 T_{y_\mathcal{I} a_\mathcal{I}}:={\rm diag}(G_{y_ia_i})+{\rm diag}(G_{y_iv_i})Q_{\mathcal{I}}{\rm diag}(G_{w_ia_i})
 \end{array}
\]
with $Q_{\mathcal{I}}:=L_{\mathcal{I}}(I-{\rm diag}(G_{w_iv_i})L_{\mathcal{I}})^{-1}$.
The entire networked control system for $\mathcal{I}$ is denoted by $\Sigma_{\mathcal{I}}$.
As with the nominal case, the resulting networked control system is also assumed to be well-posed.

\begin{figure}[t]
\centering
\includegraphics[width=.98\linewidth]{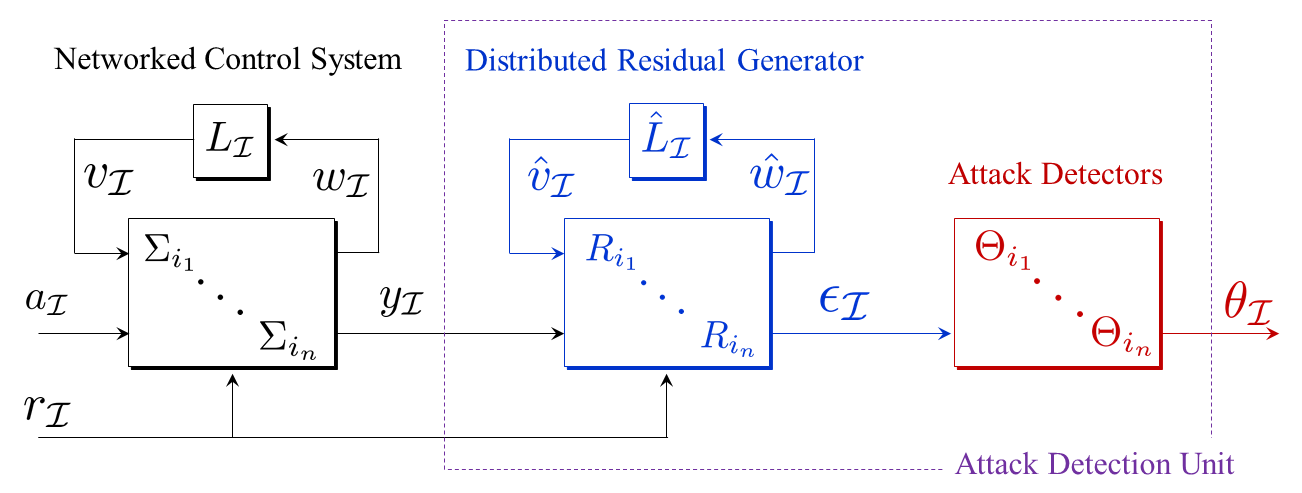}
\caption{Entire signal-flow diagram with $\mathcal{I}=\{i_1,\ldots,i_n\}$ including the networked control system to be protected, the distributed residual generator,  and the attack detectors.}
\label{fig:entire}
\end{figure}

To handle the varying network topology, the distributed residual generator is assumed to be able to modify its architecture through \emph{separation-based reconfiguration}.
The reconfigured distributed residual generator is given by~\eqref{eq:local_residual_generator} for $i\in \mathcal{I}$ with
\begin{equation}\label{eq:communication}
 \hat{v}_{\mathcal{I}}=\hat{L}_{\mathcal{I}}\hat{w}_{\mathcal{I}}
\end{equation}
where $\hat{L}_{\mathcal{I}}$ is the submatrix of $\hat{L}$ composed of its $(i,j)$th block components for $i,j\in \mathcal{I}$.
The transfer matrix $\hat{L}_{\mathcal{I}}$ can be interpreted as the resulting communication system under the separation of the local residual generators associated with the disconnected subsystems as illustrated by Fig.~\ref{fig:sys_structure}.
This reconfiguration scheme has an advantage that it can be quickly carried out by simply switching off the corresponding communication.
Moreover, a bank of pre-designed residual generators, the number of which should be enormous for large-scale systems, is not required, and hence, the proposed scheme is easy to implement.
The entire dynamics of the remaining residual generator is represented by
\begin{equation}\label{eq:res_general}
 R_{\mathcal{I}}:
 \epsilon_{\mathcal{I}} = R_{\epsilon_\mathcal{I} y_\mathcal{I}} y_{\mathcal{I}} + R_{\epsilon_\mathcal{I} r_\mathcal{I}}r_{\mathcal{I}}
\end{equation}
where $\epsilon_{\mathcal{I}}$ denotes the generated residual signal, and $R_{\epsilon_\mathcal{I} y_\mathcal{I}}$ and $R_{\epsilon_\mathcal{I} r_\mathcal{I}}$ represent the transfer matrices with respect to the subscript signals.
The entire block diagram is depicted by Fig.~\ref{fig:entire}.

\subsection{Premises for Problem Formulation}

Several premises are required for formulating the problem.
First, the networked system itself is necessarily able to operate under disconnection.
Let $\mathfrak{I}\subset 2^{\{1,\ldots,N\}}$ denote the family of all possible remaining index sets under any pre-arranged disconnection.
The following assumption is made.
\begin{assum}\label{assum:ori_sta}
The networked system $\Sigma_{\mathcal{I}}$ is internally stable for any $\mathcal{I} \in \mathfrak{I}$.
\end{assum}

Note that Assumption~\ref{assum:ori_sta} is essentially needed regardless of the choice of the attack detection unit to be designed as long as disconnection is employed as attack containment.
Note also that, although Assumption~\ref{assum:ori_sta} guarantees stability of the networked system under variable network topology, it does not guarantee stability of the attack detection unit to be designed.


In practice, arranging particular subsystems that are disconnected in compliance with each alarm, namely, choice of $\mathfrak{I}$, is included in the design process.
This arrangement, called network segmentation, is commonly adopted for suppressing attack propagation in information system security~\cite{Ackerman2017Industrial}.
Clearly, network segmentation for a networked control system should be performed such that the resulting $\mathfrak{I}$ satisfies the requirement in Assumption~\ref{assum:ori_sta}.
This procedure can be a technically difficult problem although this study does not discuss particular segmentation methods.
A brute-force approach based on numerical simulation is applicable when the networked system is not too complicated.
For large-scale systems, a passivity-based approach is promising.
When the components of the networked system are passive~\cite{Brogliat2006Dissipative} and its interaction is formed as a negative feedback,
simply choosing each passive subsystem as a part to be disconnected, which results in $\mathfrak{I}=2^{\{1,\ldots,N\}}$, leads to a proper segmentation.
Indeed, our application in Section~\ref{sec:app}, distribution network systems with distributed generation, has this remarkable property.

Subsequently, we make another premise on the attack.
We suppose a powerful attacker, who has complete knowledge of the system and abundant computational resources.
Specifically, it is assumed that the model of the networked system and the detection unit is known, that $\{r_i(t)\}_{i=1}^N,\{x_i(t)\}_{i=1}^N$ are known for any $t\geq 0$, and that the attack signal $a_i(t)$ for $i=1,\ldots,N$ can be any function generated by a causal map of $(t,\{r_i(t)\}_{i=1}^N,\{x_i(t)\}_{i=1}^N)$ that satisfies Assumption~\ref{assum:detectability} defined below.
We confine attention to the situation where the initial state is zero, i.e., $x_i(0)=0$ for $i=1,\ldots,N$.
This situation implies that the detector designer knows the steady state before the system is under possible attacks.
Note that our approach can apply even for the case where the initial state is unknown,
which is explained in Appendix~\ref{app:detection_without_initial_state}.
Under this preparation, \emph{undetectable attacks} are defined as follows~\cite{Pasqualetti2013Attack}.
\begin{defin}[Undetectable Attack]
Consider the networked system $\Sigma_{\mathcal{I}}$.
An attack $a_{\mathcal{I}}\neq 0$ is said to be \emph{undetectable with knowledge of initial state} when $y_{\mathcal{I},a_{\mathcal{I}}}=y_{\mathcal{I}}$
where $y_{\mathcal{I},a_{\mathcal{I}}}$ and $y_{\mathcal{I}}$ denote the outputs with and without the attack $a_{\mathcal{I}}$, under zero initial state, respectively.
\end{defin}

Undetectable attacks cannot be coped with using the detection framework considered in this paper.
Thus, the following assumption is made.
\begin{assum}\label{assum:detectability}
Consider the networked system $\Sigma_{\mathcal{I}}$.
There do not exist any undetectable attacks with knowledge of initial state for any $\mathcal{I} \in \mathfrak{I}$.
\end{assum}

The following lemma~\cite{Hou1998Input} characterizes existence of undetectable attacks with knowledge of initial state in the frequency domain.
\begin{lem}\label{lem:existence_undetectable_attacks}
There do \emph{not} exist any undetectable attacks with knowledge of initial state if and only if $T_{y_{\mathcal{I}} a_{\mathcal{I}}}$ is left invertible in $\mathcal{R}$.
\end{lem}

As indicated by Lemma~\ref{lem:existence_undetectable_attacks}, undetectable attacks rely on the (normal) column rank deficiency of the corresponding transfer matrix.
To eliminate the possibility of undetectable attacks, modification of system architecture is required, e.g., introduction of additional sensors.

\subsection{Disconnection-aware Attack Detection and Isolation Problems}
\label{subsec:prob}

On the above premises, we formulate the disconnection-aware attack detection problem.
\begin{prob}[Disconnection-aware Attack Detection]\label{prob:detection}
Under Assumptions~\ref{assum:ori_sta} and \ref{assum:detectability},
design local residual generators $R_i$ in~\eqref{eq:local_residual_generator} for $i=1,\ldots,N$ and a communication system $\hat{L}$ in~\eqref{eq:communication_nominal} such that $R_{\mathcal{I}}\in \mathcal{RH}_{\infty}$ and
\begin{equation}\label{eq:detectable_condition}
 \epsilon_\mathcal{I}\neq 0 \Leftrightarrow a_{\mathcal{I}}\neq 0
\end{equation}
for any $\mathcal{I} \in \mathfrak{I}$.
\end{prob}

Problem~\ref{prob:detection} is equivalent to the well-known attack detector design problem when $\mathcal{I}$ is fixed.
The difficulty arises from network topology variation caused by disconnection.
A straightforward approach is to use the Luenberger-type observer in a distributed form described by
\begin{equation}\label{eq:Luen_obs}
 R_i: \left\{
 \begin{array}{cl}
 \dot{\hat{x}}_i \hs = A_i\hat{x}_i+B_ir_i+U_i\hat{v}_i-H_i(y_i-\hat{y}_i)\\
 \hat{y}_i \hs = C_i\hat{x}_i+D_ir_i+V_i\hat{v}_i\\
 \hat{w}_i \hs = E_i\hat{x}_i+F_ir_i+W_i\hat{v}_i\\
 \epsilon_i \hs = y_i-\hat{y}_i
 \end{array}
 \right.
\end{equation}
for $i=1,\ldots,N$ with the communication system $\hat{v}=L\hat{w}$
and to determine the observer gains through linear matrix inequalities (LMIs) imposed by the tracking capability for any $\mathcal{I} \in \mathfrak{I}$.
Clearly, this LMI-based approach is inapplicable when $|\mathfrak{I}|$ is large.



As an advanced protection, we also consider attack isolation, namely, identification of the attacked subsystem.
Isolation capability under disconnection is a more serious issue than that without disconnection.
If we disconnect the wrong subsystems from the networked system, the attack cannot be eliminated, and it can potentially lead to sequential disconnection causing cascading failure.
A possible isolation method is to design the residual generator such that the $i$th residual is excited only by attacks injected into the $i$th subsystem.
The problem is formulated as follows.
\begin{prob}[Disconnection-aware Attack Isolation]\label{prob:isolation}
Under Assumptions~\ref{assum:ori_sta} and~\ref{assum:detectability}, design local residual generators $R_i$ in~\eqref{eq:local_residual_generator} for $i=1,\ldots,N$ and a communication system $\hat{L}$ in~\eqref{eq:communication_nominal} such that $R_{\mathcal{I}}\in\mathcal{RH}_{\infty}$ and
\begin{equation}\label{eq:isolation_condition}
\epsilon_i\neq 0 \Leftrightarrow a_i\neq 0
\end{equation}
for any $\mathcal{I}\in\mathfrak{I}$.
\end{prob}

As with Problem~\ref{prob:detection}, this problem can be reduced to a classic isolation problem if $\mathcal{I}$ is fixed, and the difficulty arises from the varying network topology.

\section{Proposed Disconnection-aware Residual Generator Design}
\label{sec:sol}


In this section, we propose a residual generator design method that can preserve its detection and isolation capability under separation-based reconfiguration.
Based on the proposed architecture, a solution to the formulated problem is provided.

\subsection{Design Parameters of Residual Generator}

We first review a parameterization of all residual generators in~\eqref{eq:res_general} for a fixed $\mathcal{I}$ without the constraint on the residual generator structure.
Let $(M_\mathcal{I},N_\mathcal{I})$ be a left coprime factorization of $T_{y_{\mathcal{I}}r_\mathcal{I}}$ over $\mathcal{RH}_{\infty}$~\cite[Definition~3.2]{Ding2013Model}.
Then all residual generators in~\eqref{eq:res_general} can be parameterized in the following sense~\cite[Theorem~5.3]{Ding2013Model}:
A residual generator in~\eqref{eq:res_general} satisfies $R_{\mathcal{I}}\in \mathcal{RH}_{\infty}$ and~\eqref{eq:detectable_condition} if and only if there exists a stable transfer matrix $S_\mathcal{I}\in \mathcal{RH}_{\infty}$ such that
\begin{equation}\label{eq:R_design_form}
 R_{\mathcal{I}}: \epsilon_\mathcal{I} = S_\mathcal{I}(M_\mathcal{I} y_\mathcal{I} - N_\mathcal{I} r_{\mathcal{I}})
\end{equation}
and $S_{\mathcal{I}}M_{\mathcal{I}} T_{y_{\mathcal{I}} a_{\mathcal{I}}}$ is left invertible.
Furthermore, the residual signal with the residual generator is governed by
\begin{equation}\label{eq:map_a_to_epsilon}
 \epsilon_{\mathcal{I}}= S_{\mathcal{I}}M_{\mathcal{I}} T_{y_{\mathcal{I}} a_{\mathcal{I}}}a_{\mathcal{I}}.
\end{equation}
This parameterization implies that the residual generator design problem can be reduced to finding a left coprime factorization $(M_\mathcal{I},N_\mathcal{I})$ and an appropriate $S_\mathcal{I}$ that can be realized through the structured residual generator.

In the parameterization, the pair $(M_\mathcal{I},N_\mathcal{I})$ plays the role of feedback operation, which is crucially related to response speed and robustness.
Indeed, left coprime factorization can be carried out by designing a state observer~\cite[Lemma~3.1]{Ding2013Model}.
Thus it suffices to design an observer in the form of~\eqref{eq:local_residual_generator} with~\eqref{eq:communication} for design of $(M_{\mathcal{I}},N_\mathcal{I})$.
On the other hand, $S_\mathcal{I}$ plays the role of feedforward filter, such as isolation and noise reduction.
For handling disconnection, we consider block diagonal $S_\mathcal{I}$ given by
\begin{equation}\label{eq:S_i}
 S_\mathcal{I} = {\rm diag}(S_i)_{i \in \mathcal{I}}
\end{equation}
with stable transfer matrices $S_i$ for $i \in \mathcal{I}$.
Because the block diagonal structure is preserved under disconnection, it suffices to choose appropriate $S_i$ for $i=1,\ldots,N$.

\subsection{Proposed Disconnection-aware Attack Detection}

This subsection addresses Problem~\ref{prob:detection}, namely, the detection problem.
For detection, it suffices to design only the pair $(M_{\mathcal{I}},N_{\mathcal{I}})$ because $S_{\mathcal{I}}$ can be chosen as any block diagonal left-invertible stable transfer matrix for the sake of detection (no null space).
Hence we focus only on design of $(M_{\mathcal{I}},N_{\mathcal{I}})$, or equivalently, design of a distributed observer with a given structure.
We assume that $S_i=I$ for $i=1,\ldots,N$ throughout this subsection for notational simplicity.

The crucial requirements for the observer to be designed are as follows:
\begin{itemize}
\item The observer has the structure composed of~\eqref{eq:local_residual_generator} and~\eqref{eq:communication_nominal}.
\item The structured observer preserves its tracking capability for any $\mathcal{I} \in \mathfrak{I}$.
\end{itemize}
The simplest observer that fulfills those requirements can be designed by not utilizing error feedback inside the observer, i.e.,
\begin{equation}\label{eq:naive_observer}
 \left\{
 \begin{array}{cl}
 \dot{\hat{x}}_i \hs = A_i\hat{x}_i+B_ir_i+U_i\hat{v}_i\\
 \hat{y}_i \hs = C_i\hat{x}_i+D_ir_i+V_i\hat{v}_i\\
 \hat{w}_i \hs = E_i\hat{x}_i+F_ir_i+W_i\hat{v}_i
 \end{array}
 \right.
\end{equation}
with $\hat{L}=L.$
Clearly, the first requirement is satisfied.
Moreover, since the networked control system is stable for any $\mathcal{I} \in \mathfrak{I}$ from Assumption~\ref{assum:ori_sta}, the second requirement is also satisfied.
We refer to the approach with this observer as \emph{the naive approach,} which results in
\[
 M_{\mathcal{I}}=I,\quad N_{\mathcal{I}}=T_{y_{\mathcal{I}} r_{\mathcal{I}}}
\]
for any $\mathcal{I} \in \mathfrak{I}$.
However, since the naive approach cannot move the poles of the residual generator at all, early attack detection cannot be achieved when the time constant of the attacked subsystem is large.
To design a more sophisticated attack detector, we seek for an observer different from the naive one.

\begin{figure*}[t]
\centering
\includegraphics[width=.98\linewidth]{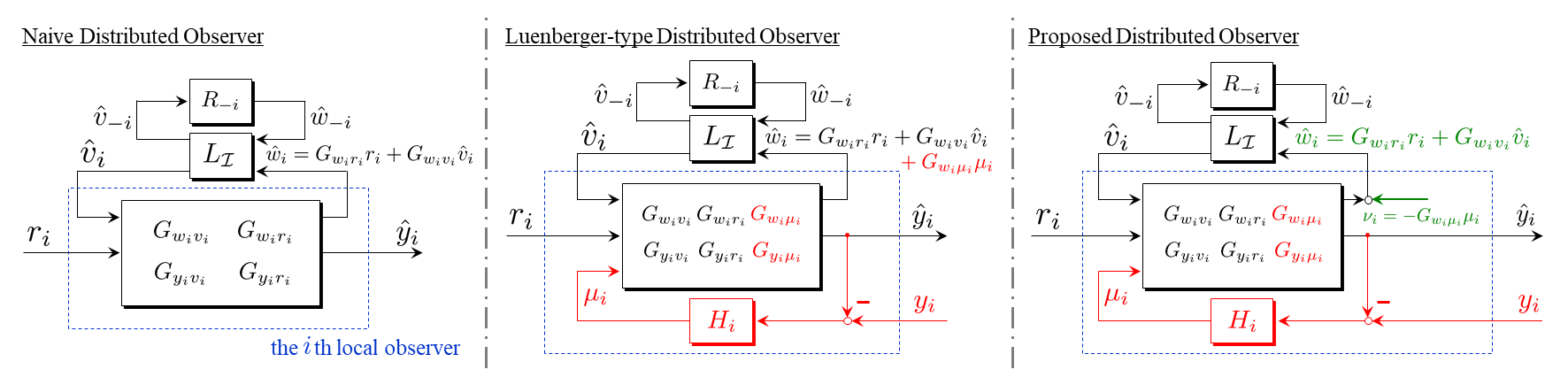}
\caption{Block diagrams of the residual generator with the naive distributed observer~\eqref{eq:naive_observer}, the Luenberger-type observer~\eqref{eq:Luen_obs}, and the proposed observer~\eqref{eq:res_retro}, where $G_{y_i\mu_i}$ is the transfer matrix from $\mu_i$ to $y_i$, $R_{-i}$ contains $R_j$ for $j\neq i,j\in \mathcal{I}$, and $\hat{v}_{-i},\hat{w}_{-i}$ are defined in a similar manner.
In the Luenberger-type observer, a feedback architecture is introduced to the naive observer through $\mu_i$.
In the proposed observer, an additional signal $\nu_i$ is injected to rectify the estimated outflowing interaction signal $\hat{w}_i$.
}
\label{fig:proposed_residual_generator}
\end{figure*}

Let us consider introducing error feedback into~\eqref{eq:naive_observer} in a decentralized manner, i.e.,
\[
 \left\{
 \begin{array}{cl}
 \dot{\hat{x}}_i \hs = A_i\hat{x}_i+B_ir_i+U_i\hat{v}_i+\mu_i\\
 \hat{y}_i \hs = C_i\hat{x}_i+D_ir_i+V_i\hat{v}_i\\
 \hat{w}_i \hs = E_i\hat{x}_i+F_ir_i+W_i\hat{v}_i
 \end{array}
 \right.
\]
with a certain feedback signal $\mu_i$.
The choice $\mu_i=H_i(y_i-\hat{y}_i)$, which results in the Luenberger-type observer~\eqref{eq:Luen_obs}, can lead to instability of the residual generator under separation-based reconfiguration even if the observer gains are determined so as to stabilize the nominal distributed observer.
Furthermore, it is difficult to find a collection of observer gains with which the distributed observer is stable for any $\mathcal{I} \in \mathfrak{I}$ as mentioned in the problem formulation.

The key observation is that the outflowing interaction signal under the error feedback is governed by
\[
 \hat{w}_i=G_{w_i r_i}r_i+G_{w_i v_i}\hat{v}_i+G_{w_i \mu_i}\mu_i
\]
where $G_{w_i \mu_i}:= E_i(sI-A_i)^{-1}$.
Because $\mu_i$ can disturb $\hat{w}_i$, the feedback interaction between the $i$th subsystem and the others are also disturbed, which can be a cause of the instability.
Thus we expect that the stability can be guaranteed by maintaining the interaction invariant under error feedback by adding an artificial input signal that rectifies $\hat{w}_i$ so as to remove the effect of $\mu_i$.

Based on this idea, we propose the following distributed observer:
\begin{equation}\label{eq:res_retro}
 \left\{
 \begin{array}{cl}
 \dot{\hat{x}}_i \hs = A_i\hat{x}_i+B_ir_i+U_i\hat{v}_i+\mu_i\\
 \hat{y}_i \hs = C_i\hat{x}_i+D_ir_i+V_i\hat{v}_i\\
 \hat{w}_i \hs = E_i\hat{x}_i+F_ir_i+W_i\hat{v}_i+\nu_i\\
 \end{array}
 \right.
\end{equation}
and
\begin{equation}\label{eq:mu_i}
 \mu_i = H_i(y_i-\hat{y}_i),\quad
 \left\{
 \begin{array}{cl}
 \dot{\hat{\chi}}_i \hs = A_i \hat{\chi}_i + \mu_i\\
 \nu_i \hs = -E_i\hat{\chi}_i,
 \end{array}
 \right.
\end{equation}
with the communication system~
\begin{equation}\label{eq:communication_same}
 \hat{L}=L.
\end{equation}
Then we have
\begin{equation}\label{eq:w_invariance}
 \begin{array}{cl}
 \hat{w}_i \hs =G_{w_i r_i}r_i+G_{w_i v_i}\hat{v}_i+G_{w_i \mu_i}\mu_i+\nu_i\\
  \hs =G_{w_i r_i}r_i+G_{w_i v_i}\hat{v}_i,
 \end{array}
\end{equation}
which is the same as that of~\eqref{eq:naive_observer}.
Thus, it is guaranteed that the stability of the residual generator is preserved under separation.
It should be emphasized that this choice does \emph{not} imply that the residual generator is not governed by those inputs.
Indeed, the estimated state $\hat{x}_i$ is affected by $\mu_i$, and hence a feedback path inside the local state observer remains even with $\nu_i$.
Block diagrams of the naive observer with~\eqref{eq:naive_observer}, the Luenberger-type observer~\eqref{eq:Luen_obs}, and the proposed observer~\eqref{eq:res_retro} are depicted in Fig.~\ref{fig:proposed_residual_generator}.
It should also be remarked that this idea originates from \emph{retrofit control}~\cite{Ishizaki2019Modularity}, which has been proposed for modular design of control systems.
The retrofit control framework is briefly reviewed in Appendix~\ref{app:retro}.

Consider the proposed residual generator composed of~\eqref{eq:res_retro} and~\eqref{eq:mu_i} with~\eqref{eq:communication_same}.
The residual generator has the distributed structure given by~\eqref{eq:local_residual_generator} and~\eqref{eq:communication_nominal}.
\begin{lem}\label{lem:res_structure}
Let the transfer matrices in~\eqref{eq:local_residual_generator} be given by
\begin{equation}\label{eq:res_specific_paramters}
 \begin{array}{lll}
 R_{\epsilon_i y_i} = M_i, & R_{\epsilon_i r_i} = -M_iG_{y_i r_i}, & R_{\epsilon_i \hat{v}_i} = -M_iG_{y_i v_i},\\
 R_{\hat{w}_i y_i} = 0, & R_{\hat{w}_i r_i}=G_{w_ir_i}, &  R_{\hat{w}_i \hat{v}_i}=G_{w_iv_i},
 \end{array}
\end{equation}
where
\begin{equation}\label{eq:Mi_def}
 M_i:=(I+G_{y_i\mu_i}H_i)^{-1}
\end{equation}
with $G_{y_i \mu_i}:= C_i(sI-A_i)^{-1}$.
Then the state-space representation of the local residual generator is given by~\eqref{eq:res_retro} with~\eqref{eq:mu_i}.
\end{lem}
\begin{IEEEproof}
By substituting~\eqref{eq:res_retro} and~\eqref{eq:mu_i} into $\epsilon_i=y_i-\hat{y}_i,$ we have
\[
 \left\{
 \begin{array}{cl}
 \dot{\hat{x}}_i \hs = A_i \hat{x}_i + B_ir_i + U_i\hat{v}_i + H_i\epsilon_i\\
 \epsilon_i \hs = y_i - C_i\hat{x}_i + D_ir_i + V_i\hat{v}_i.
 \end{array}
 \right.
\]
Thus, in the frequency domain,
\[
\epsilon_i = y_i - C_i(sI-A_i)^{-1}(B_i r_i + U_i\hat{v}_i + H_i\epsilon_i) +D_ir_i + V_i \hat{v}_i.
\]
Hence
\[
 (I+G_{y_i\mu_i}H_i)\epsilon_i = y_i - G_{y_ir_i}r_i - G_{y_iv_i}\hat{v}_i,
\]
which leads to~\eqref{eq:res_specific_paramters}.
\end{IEEEproof}

A remarkable fact is that the proposed residual generator results in a block-diagonally structured $M_{\mathcal{I}}$ given by
\[
 M_{\mathcal{I}}={\rm diag}(M_i)_{i\in\mathcal{I}}
\]
although $M_{\mathcal{I}}$ in~\eqref{eq:map_a_to_epsilon} has a dense structure in general.
The following lemma holds.
\begin{lem}\label{lem:block_diagonal_M}\marginpar{(c-3)}
Consider the proposed residual generator composed of~\eqref{eq:res_retro} and~\eqref{eq:mu_i} with~\eqref{eq:communication_same}.
Then the input-output relationships in~\eqref{eq:res_general} are given by
\begin{equation}\label{eq:Rs_entire}
 R_{\epsilon_\mathcal{I} y_\mathcal{I}}= {\rm diag}(M_i),\quad R_{\epsilon_\mathcal{I} r_\mathcal{I}}=-{\rm diag}(M_i)T_{y_\mathcal{I}r_\mathcal{I}}.
\end{equation}
Moreover, the transfer matrix from $a_{\mathcal{I}}$ to $\epsilon_{\mathcal{I}}$ is given by
\begin{equation}\label{eq:map_a_to_y_str}
 \epsilon_{\mathcal{I}}={\rm diag}(M_i) T_{y_{\mathcal{I}} a_{\mathcal{I}}} a_{\mathcal{I}}.
\end{equation}
\end{lem}
\begin{IEEEproof}
Since $R_{\hat{w}_i y_i}=0$ and $\hat{L}_\mathcal{I}=L_\mathcal{I}$, we have
\[
 \begin{array}{cl}
 \hat{v}_{\mathcal{I}} \hs = L_\mathcal{I} ( {\rm diag}(R_{\hat{w}_i r_i})r_\mathcal{I} + {\rm diag}( R_{\hat{w}_i \hat{v}_i})\hat{v}_\mathcal{I}),\\
 \hs = L_\mathcal{I} ( {\rm diag}(G_{w_ir_i})r_\mathcal{I} + {\rm diag}( G_{w_iv_i} )\hat{v}_\mathcal{I})
 \end{array}
\]
and hence $\hat{v}_{\mathcal{I}}=Q_\mathcal{I}{\rm diag}(G_{w_ir_i})r_\mathcal{I}$.
Thus Lemma~\ref{lem:res_structure} implies that
\[
 \begin{array}{cl}
 \epsilon_\mathcal{I} \hs = {\rm diag}(R_{\epsilon_i y_i})y_\mathcal{I} + {\rm diag}(R_{\epsilon_i r_i})r_\mathcal{I} + {\rm diag}(R_{\epsilon_i v_i})\hat{v}_\mathcal{I}\\
  \hs = {\rm diag}(M_i)y_\mathcal{I} - {\rm diag}(M_iG_{y_ir_i})r_\mathcal{I} - {\rm diag}(M_iG_{y_iv_i})\hat{v}_\mathcal{I}\\
  \hs = {\rm diag}(M_i)y_\mathcal{I} - {\rm diag}(M_i) T_{y_{\mathcal{I}}a_{\mathcal{I}}}r_\mathcal{I},
 \end{array}
\]
which leads to~\eqref{eq:Rs_entire}.
Moreover, by substituting~\eqref{eq:map_ra_to_y} into~\eqref{eq:Rs_entire}, we obtain~\eqref{eq:map_a_to_y_str} when $r_\mathcal{I}=0$.
\end{IEEEproof}

Lemma~\ref{lem:block_diagonal_M} proves the block diagonal structure of $M_{\mathcal{I}}$ in~\eqref{eq:map_a_to_epsilon}.
In this sense, our approach can be interpreted as a method for finding a block diagonal $M_{\mathcal{I}}$.

The above lemmas derive the following theorem, which proves detection capability of the proposed residual generator.
\begin{theorem}\label{thm:detection}
Let Assumptions~\ref{assum:ori_sta} and~\ref{assum:detectability} hold.
Consider the proposed residual generator composed of~\eqref{eq:res_retro} and~\eqref{eq:mu_i} with~\eqref{eq:communication_same}.
If $A_i-H_iC_i$ is Hurwitz for $i=1,\ldots,N$, then the residual generator is stable, i.e., $R_{\mathcal{I}}\in \mathcal{RH}_{\infty}$, and satisfies~\eqref{eq:detectable_condition} for any $\mathcal{I} \in \mathfrak{I}$.
\end{theorem}
\begin{IEEEproof}
First, we show stability of $R_{\mathcal{I}}$.
From~Lemma~\ref{lem:block_diagonal_M}, this is equivalent to stability of the transfer matrices ${\rm diag}(M_i)$ and $-{\rm diag}(M_i)T_{y_\mathcal{I}r_\mathcal{I}}$.
Assumption~\ref{assum:ori_sta} implies that $T_{y_\mathcal{I}r_\mathcal{I}}$ is stable for any $\mathcal{I}\in\mathfrak{I}$.
Hence it suffices to show stability of $M_i$ for $i=1,\ldots,N$.
Consider a state-space representation of $M_i^{-1}=I+G_{y_i\mu_i}H_i$ as
\[
\left\{
 \begin{array}{cl}
 \dot{\xi}_i \hs = A_i\xi_i + H_i\mu_i\\
 y_i \hs = C_i\xi_i + \mu_i.
 \end{array}
\right.
\]
Thus we have $\mu_i=-C_i\xi_i+y_i$.
Substituting this to the state-space representation above yields
\[
 \left\{
 \begin{array}{cl}
 \dot{\xi}_i \hs = (A_i-H_iC_i)\xi_i +H_iy_i\\
 \mu_i \hs = -C_i\xi_i + y_i,
 \end{array}
 \right.
\]
which represents a state-space representation of $M_i=(I+G_{y_i\mu_i}H_i)^{-1}$.
Because $A_i-H_iC_i$ is Hurwitz, $M_i$ is stable.

We next prove~\eqref{eq:detectable_condition}.
From Assumption~\ref{assum:detectability}, $T_{y_\mathcal{I} a_\mathcal{I}}$ is left invertible.
Moreover, $M_i$ is invertible for $i=1,\ldots,N$ from its definition~\eqref{eq:Mi_def}.
Thus ${\rm diag}(M_i) T_{y_{\mathcal{I}} a_{\mathcal{I}}}$ is left invertible.
Lemma~\ref{lem:block_diagonal_M} implies that this is the transfer matrix from $a_{\mathcal{I}}$ to $\epsilon_{\mathcal{I}}$.
Hence, from Lemma~\ref{lem:existence_undetectable_attacks},~\eqref{eq:detectable_condition} holds.
\end{IEEEproof}

Clearly, Theorem~\ref{thm:detection} provides a solution to Problem~\ref{prob:detection}.

\subsection{Proposed Disconnection-aware Attack Isolation}

This subsection addresses Problem~\ref{prob:isolation}, the attack isolation problem.
On the premise that the residual generator designed in the previous section is employed, we design $S_i$ to be an isolation filter.
When $\mathcal{I}$ is fixed, this problem can be reduced to the perfect isolation problem with unknown input decoupling~\cite[Chapter 13]{Ding2013Model}.
In the existing approach, $S_i$ is designed, for a fixed $i$, such that
\begin{equation}\label{eq:decoupling_entire}
 S_iM_iT_{y_i a_i} {\rm\ is\ left\ invertible},\ {\rm and}\ S_iM_iT_{y_i a_{-i}}=0
\end{equation}
where $T_{y_i a_i}$ and $T_{y_i a_{-i}}$ denote the transfer matrices from $a_i$ to $y_i$ and that from all attack signals except for $a_i$ to $y_i$, respectively.
Indeed, this condition is equivalent to~\eqref{eq:isolation_condition}, and there exists an isolation filter $S_i \in \mathcal{RH}_{\infty}$ that satisfies this condition if and only if
\[
 {\rm rank}\,\left[T_{y_i a_i}\ T_{y_i a_{-i}}\right] = {\rm dim}(a_i)+{\rm rank}\,T_{y_i a_{-i}}
\]
holds~\cite[Theorem~13.3]{Ding2013Model}.
Moreover, if the existence condition holds, an isolation filter can systematically be designed.
However, this existing design procedure cannot straightforwardly be applied to our problem since the condition depends on $\mathcal{I}$.

\begin{figure}[t]
\centering
\includegraphics[width = .98\linewidth]{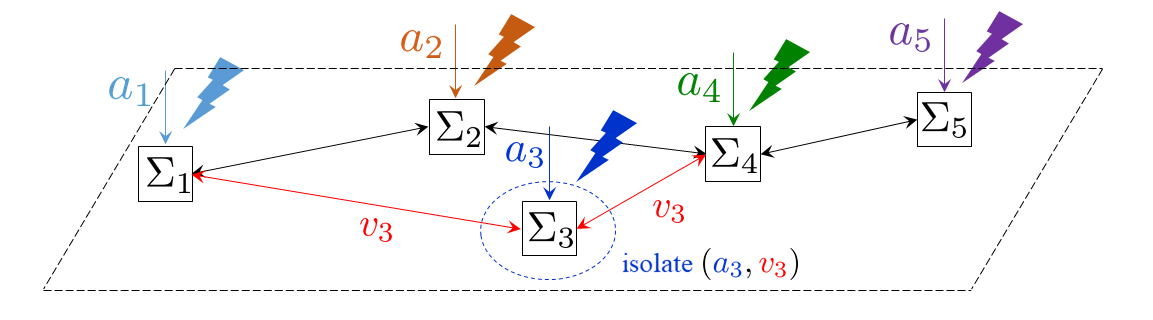}
\caption{Idea of the proposed isolation. For the third subsystem, $a_3$, the attack to be detected, and $v_3$, the inflowing interaction signal, are isolated instead of $a_3$ and the other attacks.}
\label{fig:isolation}
\end{figure}

An important observation in the proposed residual generator is that
\begin{equation}\label{eq:epsiloni_aivi}
 \begin{array}{cl}
 \epsilon_i \hs = S_iM_i(G_{y_i a_i} a_i + G_{y_i v_i}P_iQ_{\mathcal{L}}{\rm diag}(G_{w_i a_i})a_{\mathcal{I}})\\
  \hs = S_iM_i (G_{y_ia_i}a_i + G_{y_i v_i}v_i)
 \end{array}
\end{equation}
from~\eqref{eq:map_a_to_y_str}, where $P_i$ denotes the matrix that extracts $v_i$ from $v_{\mathcal{I}}$.
Hence, if $S_i$ isolates $a_i$ from $v_i$, then $a_i$ can be isolated from the other attacks regardless of $\mathcal{I}$.
This idea is illustrated by Fig.~\ref{fig:isolation}, where the attack into the third subsystem, $a_3$, and its inflowing interaction signals, $v_3$, are isolated instead of $a_3$ and the other attacks.

The following theorem describes a condition for attack isolation based on this idea.
\begin{theorem}\label{thm:isolation}
Let Assumptions~\ref{assum:ori_sta} and~\ref{assum:detectability} hold.
Consider the residual generator designed in Theorem~\ref{thm:detection}.
There exists an isolation filter $S_i$ that satisfies~\eqref{eq:isolation_condition} if
\begin{equation}\label{eq:iso_condition_ai_vi}
 {\rm rank}\,\left[G_{y_i a_i}\ G_{y_i v_i}\right] = {\rm dim}(a_i)+{\rm rank}\,G_{y_i v_i}
\end{equation}
holds.
Moreover, if $T_{v_ia_{-i}}:=P_iQ_{\mathcal{L}}G_{w_{\mathcal{I}}a_{-i}}$ is right invertible, this condition is also necessary.
\end{theorem}
\begin{IEEEproof}
\marginpar{(c-3)}The condition~\eqref{eq:iso_condition_ai_vi} is a necessary and sufficient condition for existence of a filter such that
\begin{equation}\label{eq:decoupling_local}
 S_iM_iG_{y_i a_i} {\rm\ is\ left\ invertible},\ {\rm and}\ S_iM_iG_{y_i v_i}=0
\end{equation}
from \cite[Theorem~13.3]{Ding2013Model}.
If the latter condition in~\eqref{eq:decoupling_local} holds,
\begin{equation}\label{eq:SMGyaa}
 \epsilon_i=S_iM_iG_{y_ia_i}a_i
\end{equation}
from~\eqref{eq:epsiloni_aivi}.
Hence,
\begin{equation}\label{eq:SMTyamiz}
 S_iM_iT_{y_i a_{-i}}=0
\end{equation}
holds.
Because $\epsilon_i=S_iM_i(T_{y_ia_i}a_i+T_{y_ia_{-i}}a_{-i})$, the relation~\eqref{eq:SMTyamiz} implies that $\epsilon_i=S_iM_iT_{y_ia_i}a_i.$
By comparing this equation and~\eqref{eq:SMGyaa}, we have
\[
 S_iM_iT_{y_i a_i} = S_iM_i G_{y_ia_i}.
\]
Hence, the former condition in~\eqref{eq:decoupling_local} implies that $S_iM_iT_{y_ia_i}$ is left invertible.
Thus~\eqref{eq:decoupling_entire} holds for any $\mathcal{I}\in\mathfrak{I}$, which leads to~\eqref{eq:isolation_condition} from Lemma~\ref{lem:existence_undetectable_attacks}.

For necessity, assume~\eqref{eq:isolation_condition}, which is equivalent to~\eqref{eq:decoupling_entire}.
Define $T_{v_ia_{-i}}$ such that $T_{y_ia_{-i}}=G_{y_iv_i}T_{v_ia_{-i}}.$
From the latter condition in~\eqref{eq:decoupling_entire}, we have $S_iM_iG_{y_iv_i}T_{v_ia_{-i}}=0$.
From the right invertibility of $T_{v_ia_{-i}}$, this is equivalent to $S_iM_iG_{y_i v_i}=0$, which is the latter condition in~\eqref{eq:decoupling_local}.
Now it turns out that $S_iM_iT_{y_i a_i}=S_iM_i G_{y_ia_i}$ as shown in the sufficiency part.
Thus the former condition in~\eqref{eq:decoupling_entire} leads to the former condition~\eqref{eq:decoupling_local} holds.
\end{IEEEproof}

Theorem~\ref{thm:isolation} gives a solution to Problem~\ref{prob:isolation} because $G_{y_i a_i}$ and $G_{y_i v_i}$ are independent of $\mathcal{I}$.
It should be emphasized that this beneficial property, such that attack isolation can be achieved by isolating local attacks and inflowing interaction signals, is induced from the block diagonal structure of $M_{\mathcal{I}}$.
Specific design algorithms for the isolation filter can be found in~\cite[Chapter~13]{Ding2013Model}.
Appendix~\ref{app:iso} demonstrates an intuitive design procedure using unknown input observers.

Theorem~\ref{thm:isolation} also claims necessity of the idea when the transfer matrix from $a_{-i}$ to $v_i$ is right invertible, i.e., the degree of freedom of the interaction signal is less than that of the attacks injected into the other subsystems.
This situation typically arises when the number of subsystems is much larger than that of interaction ports as illustrated by Fig.~\ref{fig:isolation}.
Hence, we expect the condition~\eqref{eq:iso_condition_ai_vi} to be necessary and sufficient for large-scale systems.

\section{Application to Low-Voltage Distribution Network with Distributed Generation}
\label{sec:app}
In this section, we treat a low-voltage distribution network with distributed generation.
In distribution grids, disconnecting subsystems, which are given as distributed generations with inverters owned by customers, is related to the problem of load shedding.
Load shedding is an operation of islanding loads by opening distribution circuit breakers in order to balance supply and demand.
It has been reported that unexplained activation of inappropriate load shedding program led to a huge outage in 2007~\cite{Going2016Smith}.
The accident demonstrates the risk that should be accounted for in distribution networks although it is not intentionally caused.
We propose an algorithm that can appropriately shed loads as a counteraction against malicious attacks injected into inverters as an application of the method in Section~\ref{sec:sol}.
For general security issues in power grids, see~\cite{Sridhar2012Cyber,Li2017Cybersecurity}.

\subsection{Distribution Network Model}
\label{subsec:grid}

We first provide a mathematical model of low-voltage distribution networks with distributed generation.
Consider a rooted tree graph $\mathcal{G}=(\mathcal{N},\mathcal{B})$ that represents a radial distribution network where $\mathcal{N}$ and $\mathcal{B}\subset \mathcal{N}\times \mathcal{N}$ denote the sets of buses and branches, respectively.
Let $\mathcal{N}_{\rm DG}\subset \mathcal{N}$ denote the index set of the distributed generation buses.
The voltage magnitude at each bus is denoted by ${\rm v}_k$ for $k\in \mathcal{N}$.
The complex power flow from the $l$th bus to the $k$th bus is denoted by ${\rm S}_{lk}={\rm P}_{lk}+{\rm j}{\rm Q}_{lk}$ for $(l,k) \in \mathcal{B}$ where ${\rm j}$ is the imaginary unit.
Lines from the $l$th bus to the $k$th bus have an impedance ${\rm Z}_{lk}={\rm R}_{lk}+{\rm j}{\rm X}_{lk}$ for $(l,k) \in \mathcal{B}$.
The voltage magnitude at the substation bus (the root node) is assumed to be a constant $\overline{{\rm v}}_0.$
For interaction among those physical quantities, we employ the LinDistFlow model~\cite{Baran1989Optimal}, which is commonly used for representing power flow and voltage magnitude drop in radial networks.
The power flow equation at each bus without distributed generation is given by
\[
 \textstyle{{\rm S}_{lk} = \sum_{m \in \mathcal{N}^{\rm out}_k} {\rm S}_{km},\quad l \in \mathcal{N}^{\rm in}_k, k\in \mathcal{N}}\setminus\mathcal{N}_{\rm DG}
\]
where $\mathcal{N}^{\rm in}_k$ and $\mathcal{N}^{\rm out}_k$ represent the inflowing buses to and the outflowing buses from the $k$th bus, respectively.
Note that the power losses in lines are assumed to be zero in this model.
The voltage drop equation at each branch is given by
\[
 {\rm v}_l^2-{\rm v}_k^2=2{\rm f}({\rm S}_{lk}),\quad (l,k)\in \mathcal{B}
\]
with ${\rm f}({\rm S}_{lk}):={\rm R}_{lk}{\rm P}_{lk}+{\rm X}_{lk}{\rm Q}_{lk}$.

We next give a model of distributed generation with inverters.
The operation of the inverter is to regulate the corresponding voltage magnitude by generating reactive power.
As the inverter dynamics, we employ the first-order model used in~\cite{Chong2019Local}, where the input signal is the deviation of squared voltages between the reference value and the actual value at the bus and the output signal is the generated reactive power.
Its dynamics is represented by
\begin{equation}\label{eq:inverter}
\left\{
 \begin{array}{cl}
 \dot{{\rm q}}_k \hs =-(1/{\rm T}_k){\rm q}_k+({\rm K}_k/{\rm T}_k)(\overline{{\rm v}}_k^2-{\rm v}_k^2)\\
 {\rm S}_{{\rm DG}_k} \hs = {\rm p}^{\rm g}_k-{\rm p}^{\rm c}_k+{\rm j}({\rm q}_k-{\rm q}^{\rm c}_k)\\
 {\rm S}_{lk} \hs =\sum_{m\in \mathcal{N}_k^{\rm out}}{\rm S}_{km} - {\rm S}_{{\rm DG}_k}, \ l\in \mathcal{N}^{\rm in}_k
 \end{array}
 \right.,\ k\in \mathcal{N}_{\rm DG}
\end{equation}
where ${\rm q}_k$ is the generated reactive power, $\overline{{\rm v}}_k$ is the reference voltage magnitude, ${\rm S}_{{\rm DG}_k}$ is the generated complex power, ${\rm p}^{\rm g}_k$ is a fixed active power generated by the distributed generation, ${\rm p}^{\rm c}_k$ is a fixed active power consumed by the customer, ${\rm q}^{\rm c}_k$ is a fixed consumed reactive power, ${\rm T}_k>0$ is the time constant of the inverter, and ${\rm K}_k\geq 0$ is a droop gain.
We suppose that the reference voltage magnitudes $\overline{{\rm v}}_k$ are identically set to the substation's reference voltage magnitude $\overline{{\rm v}}_0$.

\subsection{Representation as Networked System}

First of all, we represent the dynamics of the distribution network in the form of~\eqref{eq:ori_sys} and~\eqref{eq:ori_interaction_nominal}.
Suppose that the distribution network is partitioned into multiple segments, each of which is a collection of buses.
Let $\mathcal{G}_i=(\mathcal{N}_i,\mathcal{B}_i)$ be the subgraph corresponding to the $i$th subnetwork.
The subnetwork can be taken to be any component in the grid such as a microgrid or a single generator.
The set of the distributed generation buses in the $i$th subnetwork is denoted by $\mathcal{N}_{{\rm DG},i}$.

\begin{figure}[t]
\centering
\includegraphics[width = .98\linewidth]{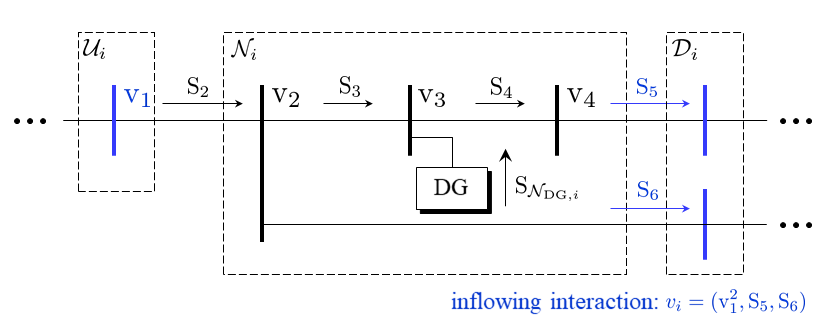}
\caption{Example of a subnetwork in a radial distribution network.
The inflowing interaction is taken to be the stacked vector of the squared voltage magnitude at the upstream bus and the power flow to the downstream buses.
}
\label{fig:ex_grid}
\end{figure}

We exemplify a construction procedure of a well-defined subsystem $\Sigma_i$ in~\eqref{eq:ori_sys} from the given subnetwork.
Consider the particular subnetwork illustrated by Fig.~\ref{fig:ex_grid}.
The $i$th subnetwork is composed of the buses $\mathcal{N}_i=\{2,3,4\}$.
Because the distribution network is radial, there exists parent buses of the subnetwork.
We call these buses the upstream buses of $\mathcal{N}_i$ denoted by $\mathcal{U}_i=\{1\}$.
On the other hand, there are buses to which power flows from the subnetwork.
We call those buses the downstream buses of $\mathcal{N}_i$ denoted by $\mathcal{D}_i=\{5,6\}$.
The notation in Fig.~\ref{fig:ex_grid} is used for the following discussion.

It suffices to find signals that determine the power flows ${\rm S}_{\mathcal{N}_i}$ and the squared voltage magnitudes ${\rm v}^2_{\mathcal{N}_i}$ for obtaining a well-defined input-output mapping.
Consider the power flow equation with respect to $\mathcal{N}_i$ given by
\[
\arraycolsep=2pt
 \underbrace{\left[
 \begin{array}{ccc}
 1 & -1 & 0\\
 0 & 1 & -1\\
 0 & 0 & 1
 \end{array}
 \right]}_{=:-{\rm M}_{\mathcal{N}_i}}
 \underbrace{\left[
 \begin{array}{c}
 {\rm S}_2\\
 {\rm S}_3\\
 {\rm S}_4
 \end{array}
 \right]}_{{\rm S}_{\mathcal{N}_i}}+
 \underbrace{\left[
 \begin{array}{cc}
 0 & -1\\
 0 & 0\\
 -1 & 0
 \end{array}
 \right]}_{=:-{\rm M}_{\mathcal{N}_i \mathcal{D}_i}}
 \underbrace{\left[
 \begin{array}{c}
 {\rm S}_5\\
 {\rm S}_6
 \end{array}
 \right]}_{{\rm S}_{\mathcal{D}_i}}
 +
 \hspace{-2mm}
 \underbrace{\left[
 \begin{array}{c}
 0\\
 1\\
 0
 \end{array}
 \right]}_{=: {\rm M}_{{\rm DG},i}}
 \hspace{-2mm}
 {\rm S}_{\mathcal{N}_{{\rm DG},i}}=0.
\]
Because ${\rm M}_{\mathcal{N}_i}$ is nonsingular, ${\rm S}_{\mathcal{N}_i}$ is uniquely determined when ${\rm S}_{\mathcal{D}_i}$ and ${\rm S}_{\mathcal{N}_{{\rm DG}},i}$ are given.
Consider also the voltage drop equation with respect to $\mathcal{N}_i$ given by
\[
\arraycolsep=2pt
 \underbrace{\left[
 \begin{array}{ccc}
 -1 & 0 & 0\\
 1 & -1 & \\
 0 & 1 & -1
 \end{array}
 \right]}_{{\rm M}^{\sf T}_{\mathcal{N}_i}}
 \underbrace{\left[
 \begin{array}{c}
 {\rm v}^2_2\\
 {\rm v}^2_3\\
 {\rm v}^2_4
 \end{array}
 \right]}_{{\rm v}^2_{\mathcal{N}_i}}
 +
 \underbrace{\left[
 \begin{array}{c}
 1 \\
 0 \\
 0
 \end{array}
 \right]}_{=:{\rm M}^{\sf T}_{\mathcal{N}_i \mathcal{U}_i}}
 \underbrace{{\rm v}^2_1}_{{\rm v}^2_{\mathcal{U}_i}}
 = \left[
 \begin{array}{c}
 2{\rm f}({\rm S}_2)\\
 2{\rm f}({\rm S}_3)\\
 2{\rm f}({\rm S}_4)
 \end{array}
 \right].
\]
Similarly, ${\rm v}^2_{\mathcal{N}_i}$ is determined from ${\rm v}^2_{\mathcal{U}_i}$ and ${\rm S}_{\mathcal{N}_i}$.
Thus, we take the inflowing interaction signal in~\eqref{eq:ori_sys} as
\begin{equation}\label{eq:inflow_interaction_DN}
 v_i := [{\rm v}_{\mathcal{U}_i}^2\ {\rm P}_{\mathcal{D}_i}^{\sf T}\ {\rm Q}_{\mathcal{D}_i}^{\sf T}]^{\sf T},
\end{equation}
which is $[{\rm v}_1^2\ {\rm P}_{5}\ {\rm Q}_{5}\ {\rm P}_{6}\ {\rm Q}_{6}]^{\sf T}$ in the example in Fig.~\ref{fig:ex_grid}.
The state, the outflowing interaction signal, and the reference signal are taken as
\begin{equation}\label{eq:state_reference_DN}
\begin{array}{l}
 x_i := {\rm q}_{\mathcal{N}_{{\rm DG},i}},\quad w_i := [{\rm v}_{\mathcal{N}_i}^{2 {\sf T}}\ {\rm P}_{\mathcal{N}_i}^{\sf T}\ {\rm Q}_{\mathcal{N}_i}^{\sf T}]^{\sf T},\\
 r_i:=\left[\ol{{\rm v}}^2_0\ \ol{{\rm v}}^{2 {\sf T}}_{\mathcal{N}_{{\rm DG},i}}\ {\rm p}^{{\rm g} {\sf T}}_{\mathcal{N}_{{\rm DG},i}}\ {\rm p}^{{\rm c}{\sf T}}_{\mathcal{N}_{{\rm DG},i}}\ {\rm q}^{{\rm c}{\sf T}}_{\mathcal{N}_{{\rm DG},i}}\right]^{\sf T}.
\end{array}
\end{equation}
Then we obtain the system representation in the form~\eqref{eq:ori_sys}.

The following proposition guarantees that the networked system induced by this choice is well-defined as an input-output map in general.
\begin{prop}\label{prop:dist_net_representation}
Consider a radial distribution network with a given partition.
Take the inflowing interaction signal as~\eqref{eq:inflow_interaction_DN} and the other signals as~\eqref{eq:state_reference_DN}.
Then the subsystems $\Sigma_i$ in~\eqref{eq:ori_sys} and their interaction~\eqref{eq:ori_interaction} are well-defined for any $\mathcal{I}\in 2^{\{1,\ldots,N\}}$.
Moreover, the entire networked system is well-posed.
\end{prop}
\begin{IEEEproof}
Let ${\rm P}$ denote the stacked vector of all active power flows and other vectors are similarly denoted.
From the matrix form of the LinDistFlow model~\cite{Zhu2016Fast}, the entire power flow is given by
\[
\begin{array}{cl}
 -{\rm M}{\rm P}+{\rm M}_{{\rm DG}}{\rm P}_{\mathcal{N}_{\rm DG}}\hs=0,\\
 -{\rm M}{\rm Q}+{\rm M}_{{\rm DG}}{\rm Q}_{\mathcal{N}_{\rm DG}}\hs=0
\end{array}
\]
and the voltage drop is given by
\[
 {\rm M}^{\sf T}{\rm v}^2+{\rm m}\overline{{\rm v}}^2_0=2{\rm D}_{\rm R}{\rm P}+2{\rm D}_{\rm X}{\rm Q}
\]
with a matrix ${\rm M}_{\rm DG}$ where $[{\rm m}\ {\rm M}^{\sf T}]^{\sf T}$ denotes the graph incidence matrix and ${\rm D}_{\rm R},{\rm D}_{\rm X}$ represent the diagonal matrices whose components are corresponding resistances and the reactances, respectively.
Thus, subtracting the relevant rows, we obtain the LinDistFlow model for the $i$th subnetwork as
\[
\begin{array}{cl}
 -{\rm M}_{\mathcal{N}_i}{\rm P}_{\mathcal{N}_i}-{\rm M}_{\mathcal{N}_i\mathcal{D}_i}{\rm P}_{\mathcal{D}_i}+{\rm M}_{{\rm DG},i}P_{\mathcal{N}_{{\rm DG},i}}\hs =0,\\
 -{\rm M}_{\mathcal{N}_i}{\rm Q}_{\mathcal{N}_i}-{\rm M}_{\mathcal{N}_i\mathcal{D}_i}{\rm Q}_{\mathcal{D}_i}+{\rm M}_{{\rm DG},i}Q_{\mathcal{N}_{{\rm DG},i}}\hs =0\\
\end{array}
\]
and
\[
 {\rm M}^{\sf T}_{\mathcal{N}_i}{\rm v}_{\mathcal{N}_i}^2 + {\rm M}^{\sf T}_{\mathcal{N}_i\mathcal{U}_i}{\rm v}_{\mathcal{U}_i}^2+{\rm m}_{\mathcal{N}_i}\overline{\rm v}_0^2=2{\rm D}_{{\rm R},i}{\rm P}_{\mathcal{N}_i}+2{\rm D}_{{\rm X},i}{\rm Q}_{\mathcal{N}_i},
\]
with the subtracted matrices.
Because ${\rm M}_{\mathcal{N}_i}$ is nonsingular~\cite{Zhu2016Fast},
${\rm P}_{\mathcal{N}_i},{\rm Q}_{\mathcal{N}_i},{\rm v}^2_{\mathcal{N}_i}$ are uniquely determined from $x_i,r_i,v_i$.
From~\eqref{eq:inverter}, $x_i$ is uniquely determined from $r_i$ and $v_i$ with any initial state.
Since the inverter dynamics from ${\rm v}^2_{k}$ to ${\rm q}_k$ is strictly proper, the feedback system composed of $x_i$ and ${\rm v}^2_{\mathcal{N}_{{\rm DG}_i}}$ is well-posed.
Thus the subsystem $\Sigma_i$ is well-defined.

We show that the interaction is also well-defined.
Let $\overline{\mathcal{N}}:=\bigcup_i \mathcal{N}_i$ and $\mathcal{M}:=\mathcal{N}\setminus \overline{\mathcal{N}}$.
Because $\mathcal{N}_{\rm DG} \subset \overline{\mathcal{N}}$, we have
\[
 \begin{array}{cl}
 -{\rm M}_{\mathcal{M}} {\rm P}_{\mathcal{M}} - {\rm M}_{\mathcal{M} \overline{\mathcal{N}}}P_{\overline{\mathcal{N}}}\hs=0,\\
 -{\rm M}_{\mathcal{M}} {\rm Q}_{\mathcal{M}} - {\rm M}_{\mathcal{M} \overline{\mathcal{N}}}Q_{\overline{\mathcal{N}}}\hs=0
 \end{array}
\]
and
\[
 {\rm M}_{\mathcal{M}}^{\sf T} {\rm v}_{\mathcal{M}}^2+{\rm M}_{\mathcal{M} \overline{\mathcal{N}}}^{\sf T} {\rm v}_{\overline{\mathcal{N}}}^2+{\rm m}_{\mathcal{M}} \overline{{\rm v}}_0^2 = 2{\rm D}_{{\rm R}\mathcal{M}}{\rm P}_{\mathcal{M}}+2{\rm D}_{{\rm X} \mathcal{M}}{\rm Q}_{\mathcal{M}}.
\]
Because ${\rm M}_{\mathcal{M}}$ is nonsingular, $P_\mathcal{M},Q_\mathcal{M},{\rm v}^2_\mathcal{M}$ are determined from $w_\mathcal{I}$ and the reference signal.
Those vectors contain $v_{\mathcal{I}}$.
Thus, the entire system is well-posed.
\end{IEEEproof}

The most important feature of this representation is that the interaction structure has the same network topology as that of the original graph.
Hence, we can naturally employ this network, which is typically sparse, for communication topology of our distributed residual generator.

\subsection{Stability of Distribution Network under Disconnection}

We show that any distribution network fulfills Assumption~\ref{assum:ori_sta} as long as a collection of buses forms a subsystem.
Because reference signals are irrelevant to stability from linearity of the system, the reference voltage magnitudes, the generated/consumed active powers, and the consumed reactive powers are assumed to be zero in this subsection.
The following lemma~\cite{Farivar2013Equilibrium} provides another representation of the distribution network used for stability analysis.
\begin{lem}\label{lem:map_q_to_v}
Assume that all exogenous reference signals are zero.
The input-output map from ${\rm q}_{\mathcal{N}_{\rm DG}}$ to ${\rm v}^2_{\mathcal{N}_{\rm DG}}$ for any radial distribution network can be represented by ${\rm v}^2_{\mathcal{N}_{\rm DG}}={\rm X}{\rm q}_{\mathcal{N}_{\rm DG}}$
with a positive definite matrix ${\rm X}$.
\end{lem}

Lemma~\ref{lem:map_q_to_v} indicates that the state behavior can be represented by
\begin{equation}\label{eq:inv_passive}
 \dot{{\rm q}}_k=-(1/{\rm T}_k){\rm q}_k+({\rm K}_k/{\rm T}_k)(-{\rm v}^2_k),\quad k\in \mathcal{N}_{\rm DG}
\end{equation}
and
\begin{equation}\label{eq:X_passive}
 {\rm v}^2_{\mathcal{N}_{\rm DG}}={\rm X}{\rm q}_{\mathcal{N}_{\rm DG}}.
\end{equation}
Because those systems are strictly positive real and the feedback is negative,
the entire system is stable.
Thus stability of the network is preserved under any disconnection of buses.
The following theorem holds.
\begin{theorem}\label{thm:dist_stability}
Consider a radial distribution network with distributed generation whose mathematical model is given by the LinDistFlow model and the inverter dynamics~\eqref{eq:inverter}.
This system is internally stable under any disconnection of buses.
\end{theorem}
\begin{IEEEproof}
Because~\eqref{eq:inv_passive} is a stable single-input single-output system and the signs of the coefficients for the input and output signals are positive, the inverter dynamics is strictly positive real~\cite[Definition~2.42]{Brogliat2006Dissipative}.
Hence, the diagonal system composed of the inverters is also strictly positive real.
On the other hand, because ${\rm X}$ is positive definite from Lemma~\ref{lem:map_q_to_v}, the map from ${\rm q}_{\mathcal{N}_{\rm DG}}$ to ${\rm v}_{\mathcal{N}_{\rm DG}}$ is also strictly positive real.
Because their interaction forms a negative feedback, the entire system is internally stable.
The internal stability is guaranteed for any network topology as long as the network is radial.
Because the radial property is preserved under disconnection of buses, the resulting distribution network is also internally stable.
\end{IEEEproof}

Theorem~\ref{thm:dist_stability} claims that distribution network systems with inverter-based distributed generation modeled by~\eqref{eq:inverter} satisfy the condition of Assumption~\ref{assum:ori_sta} by letting each element of $\mathfrak{I}$ contain a collection of buses.

\emph{Remark:}
The system representation~\eqref{eq:inv_passive} and~\eqref{eq:X_passive}, which is used for proving stability, is inadequate for our distributed residual generator design.
In~\eqref{eq:X_passive}, the interaction matrix $L_{\mathcal{I}}$ in~\eqref{eq:ori_interaction} is given as ${\rm X}$, which is a dense matrix in general.
Thus, this system representation means that the communication structure among the local residual generators to be designed becomes also dense although the graph under a typical distribution network is sparse.
It should also be remarked that, it is unclear that the system representation in Proposition~\ref{prop:dist_net_representation} has the passivity property.
Indeed, the dimensions of $v_i$ and $w_i$ can be different, and hence the supply rate cannot even be taken as long as the interactions signals are not reduced further.
Thus, passivity cannot straightforwardly be utilized for residual generator design with separation-based reconfiguration in this application.

In summary, it has been shown that this application satisfies the crucial assumption in our framework.
In the next section, we will illustrate the proposed residual generator design and its practical impacts by means of this example.

\section{Simulation with CIGRE Benchmark Model}
\label{sec:sim}

\subsection{Simulation Setup}
\begin{figure}[t]
\centering
\includegraphics[width = .98\linewidth]{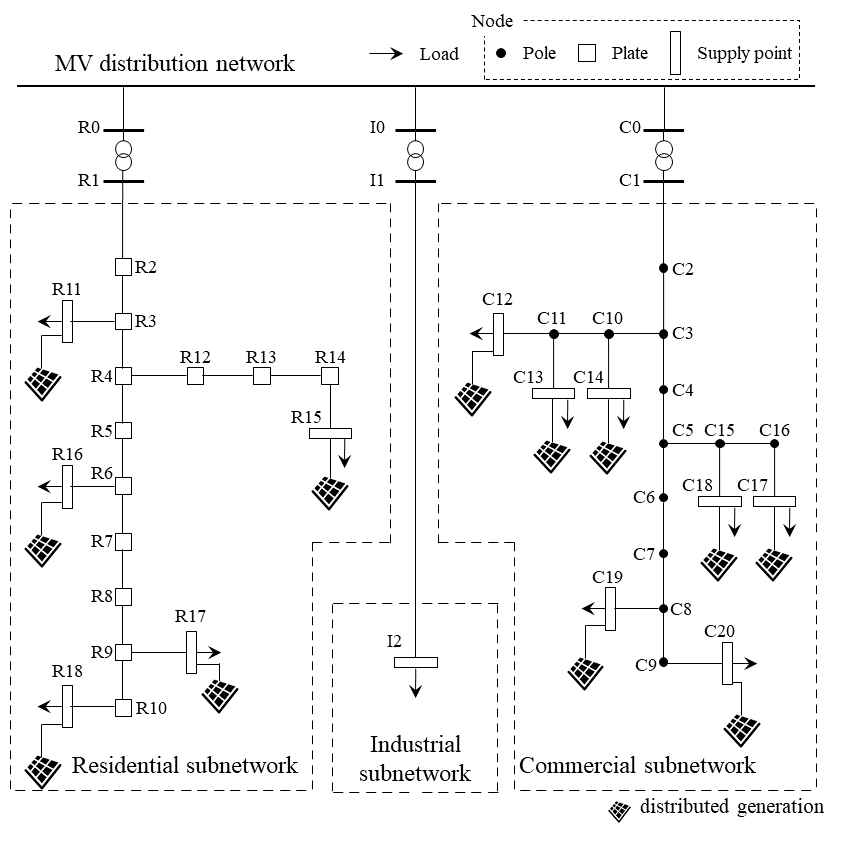}
\caption{Schematic diagram of the CIGRE benchmark model of a European low-voltage power distribution network with distributed generation.}
\label{fig:CIGRE}
\end{figure}

We provide numerical results for the CIGRE benchmark model of a European low-voltage power distribution network~\cite{Strunz2014Benchmark} with distributed generation, whose infrastructure is illustrated in Fig.~\ref{fig:CIGRE}.
In particular, we treat the residential subnetwork.
The line impedance is computed from the dataset in~\cite{Strunz2014Benchmark} on the premise that the network is balanced.
We suppose that the customer at every load has distributed generation with an inverter.
The generated/consumed active power and the consumed reactive power of the distributed generation are given in Table~\ref{tab:paras}.
The time constant of the inverters and the droop gains are uniformly given by ${\rm T}_k=2$ s and ${\rm K}_k=2$, respectively.

\begin{table}[t]
\centering
\caption{Active/reactive powers in the residential subnetwork}
\begin{tabular}{l|rrrrr}
bus & R11 & R15 & R16 & R17 & R18\\ \hline
${\rm p}^{\rm g}_k$ (W) & 3500 & 5500 & 4000 & 4500 & 3000\\
${\rm p}^{\rm c}_k$ (W) & 2295 & 5440 & 5440 & 2295 & 2720\\
${\rm q}^{\rm c}_k$ (VAr) & 300 & 960 & 480 & 600 & 400
\end{tabular}
\label{tab:paras}
\end{table}

\begin{table}[t]
\centering
\caption{Partition of the residential subnetwork}
\begin{tabular}{c|l}
 & buses composed of the subsystem\\ \hline
$\Sigma_1$ & R9, R10, R17, R18\\
$\Sigma_2$ & R2, R3, R4, R5, R6, R7, R8, R11, R12, R13, R14, R15, R16\\
\end{tabular}
\label{tab:subsystems_residential}
\end{table}

The partition of the distribution network is given in Table~\ref{tab:subsystems_residential}.
This system satisfies Assumption~\ref{assum:ori_sta} from the discussion in the previous subsection.
We suppose that the generated reactive powers of the inverters can be measured by the residual generator.
The error feedback gain inside the $i$th local residual generator $H_i$ is determined through the linear quadratic regulator (LQR) design method.
The detector $\Theta_i$ is designed such that
\[
 \Theta_i: \theta_i(t) =\left\{
 \begin{array}{ll}
 {\rm alarm}, & {\rm if\ }\|\epsilon_i(t)\|>\gamma_i,\\
 {\rm no\ alarm}, & {\rm otherwise,}
 \end{array}
 \right.
\]
where $\|\cdot\|$ denotes the Euclidean norm and $\gamma_i$ represents a prescribed threshold.
The threshold is set to avoid false alarms because of noise and model errors.

As a threat model targeting the distribution network, a voltage reference attack treated in~\cite{Teixeira2015Voltage,Isozaki2016Detection} is supposed to be injected into the distributed generation at the $k_0$th bus for a fixed $k_0 \in \mathcal{N}_{\rm DG}$.
The effect of the attack is modeled as a simple step function beginning at $t_0$ s.
Accordingly, the fabricated squared reference voltage magnitude is represented by
\[
 \overline{\rm v}_{k_0}^2(t) = \left\{
 \begin{array}{cl}
 \overline{{\rm v}}_0 ^2, & {\rm if}\ t\leq t_0, \\
 \overline{a}, & {\rm otherwise}
 \end{array}
 \right.
\]
where $\overline{a}$ is a positive scalar value that represents the amplitude of the attack.
The attack is injected into the bus R18 starting at $t_0=1$.
Accordingly, the detection threshold $\gamma_i$ is determined to be $\overline{a} \alpha_i$ where $\alpha_i$ is the Euclidean norm of the DC (direct current) gain from the attack input port to the residual relevant to the attacked subsystem.
The objective of the attack is to amplify deviation of the voltages at all distributed generator buses from the reference value.

We compare the proposed distributed attack detector with the naive distributed attack detector.
Three distributed state observers are designed under different weights for the LQR design.
The state and input weights are given as $Q=qI$ and $R=rI$ with the identity matrix $I$ whose dimension is compatible with the corresponding signals, where
\[
 q\in\{1,10\}
\]
and $r=1$, respectively.
We refer to the resulting gains as ``low gain'' and ``high gain'' respectively.

\subsection{Simulation Results}
\begin{figure*}[t]
\centering
\subfloat[][]{
\includegraphics[width=.33\linewidth]{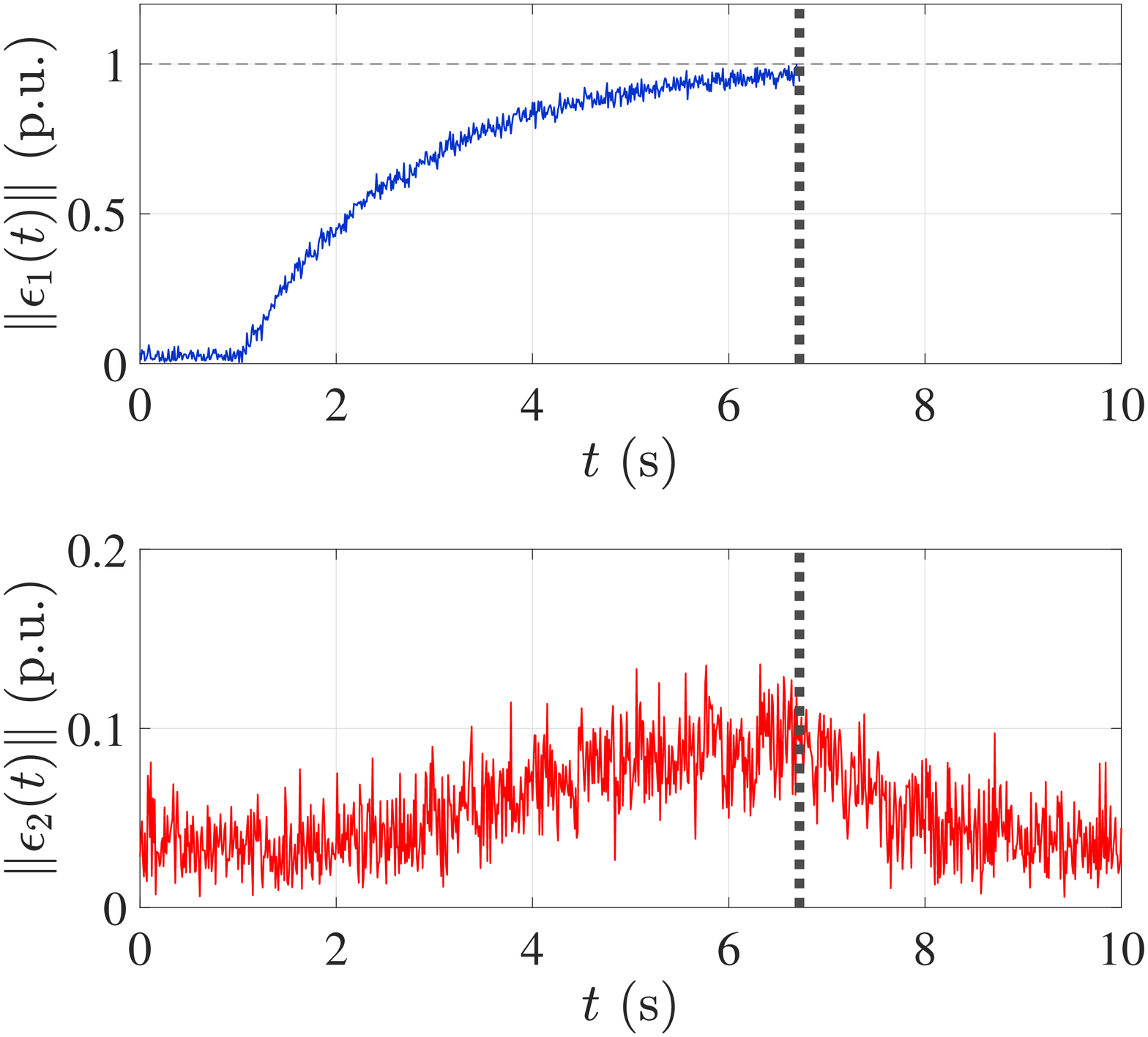}\label{subfig:err1_naive}
}
\subfloat[][]{
\includegraphics[width=.33\linewidth]{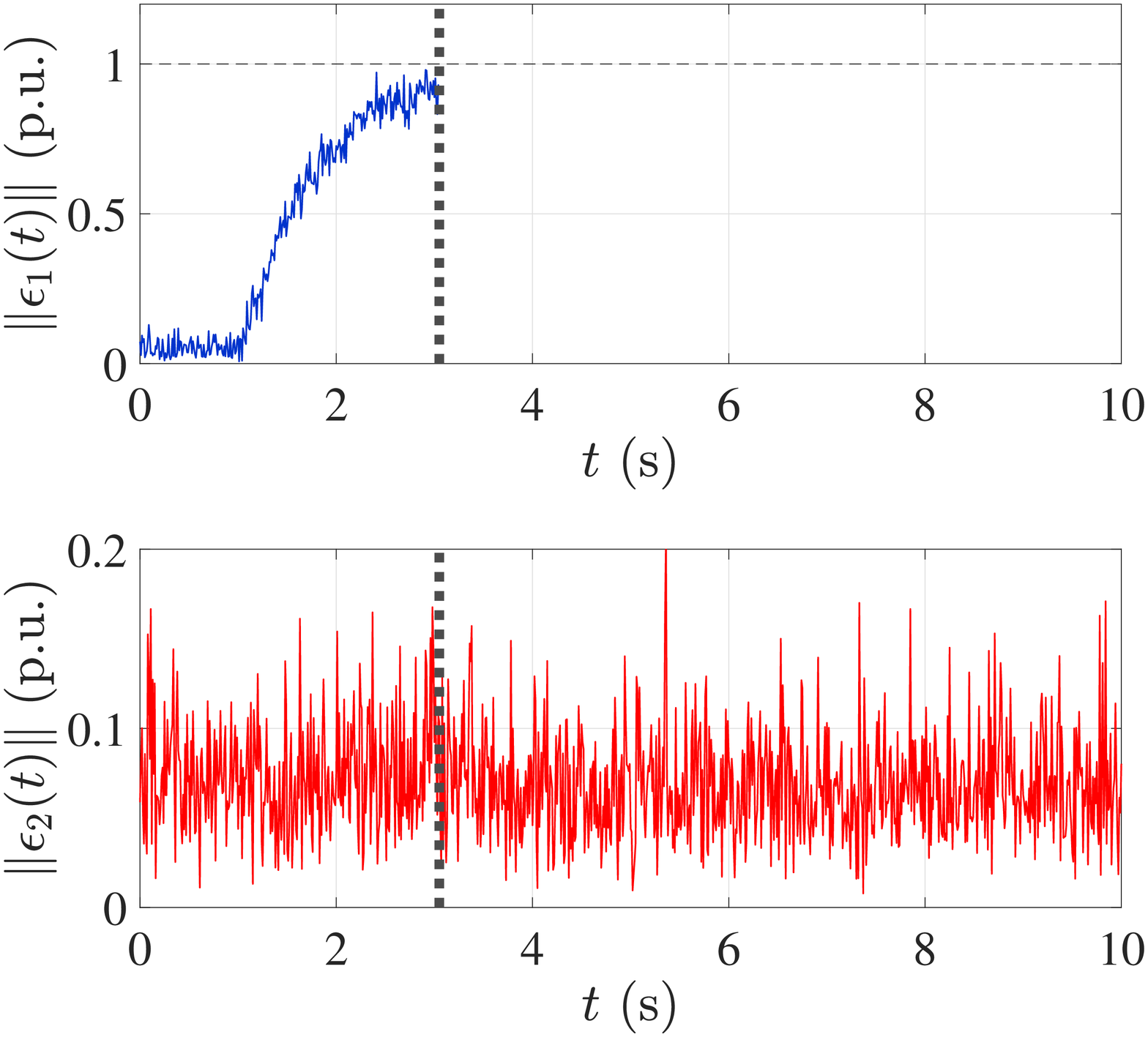}\label{subfig:err1_low}
}
\subfloat[][]{
\includegraphics[width=.33\linewidth]{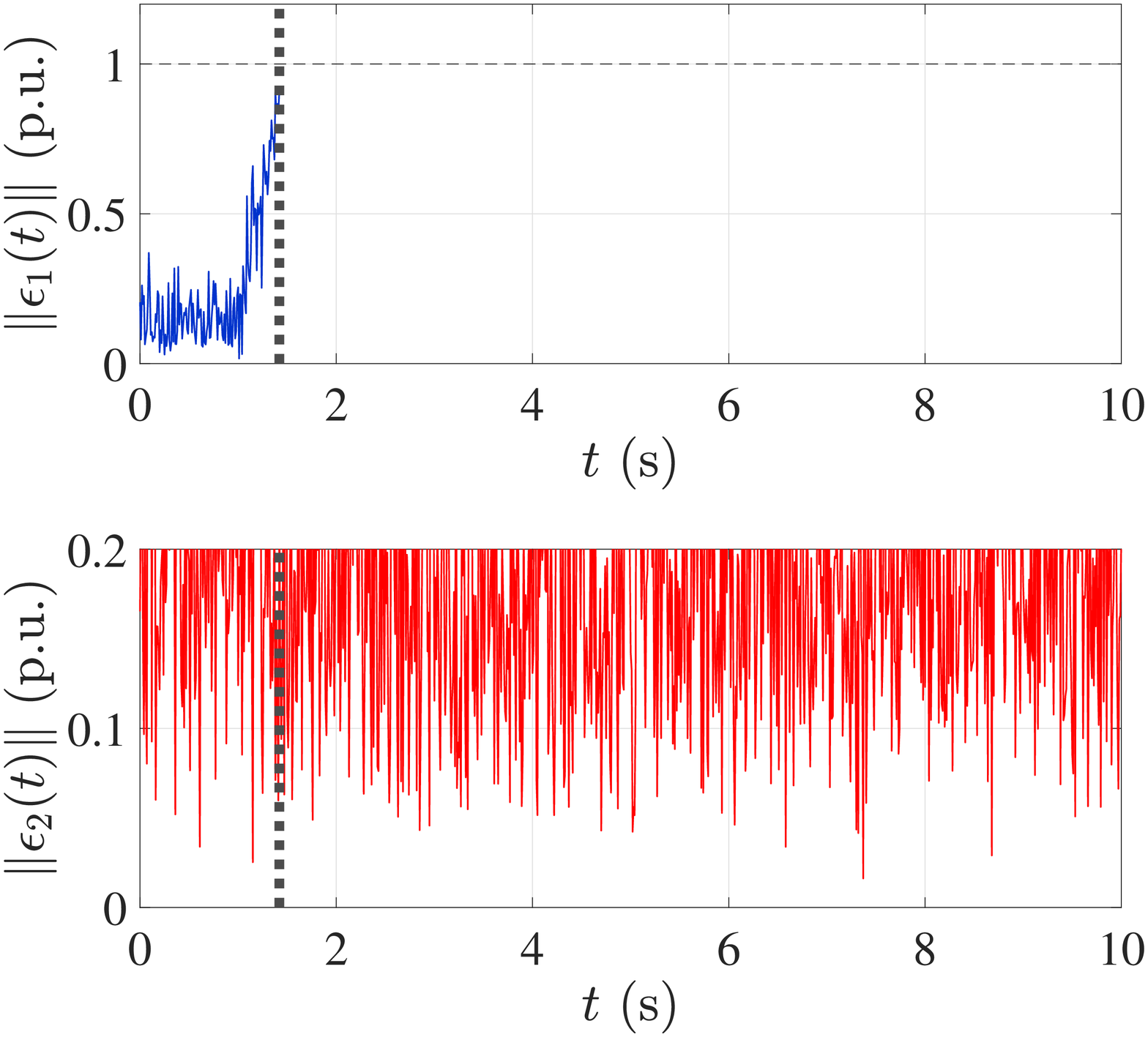}\label{subfig:err1_high}
}
\caption{Time series of $\|\epsilon_i(t)\|$ in per unit.
The top and bottom subfigures correspond to the attacked subsystem $\Sigma_1$ and the non-attacked subsystems $\Sigma_2$, respectively.
The horizontal broken lines at the top subfigures depict the prescribed detection threshold.
The vertical dotted lines depict the detection and disconnection time instants.
The signals of the separated detector are not depicted after the disconnection.
(a): Naive approach.
(b): Proposed approach with the low gain.
(c): Proposed approach with the high gain.
}
\label{graph:error_scenario1}
\end{figure*}

\begin{figure*}[t]
\centering
\subfloat[][]{
\includegraphics[width=.33\linewidth]{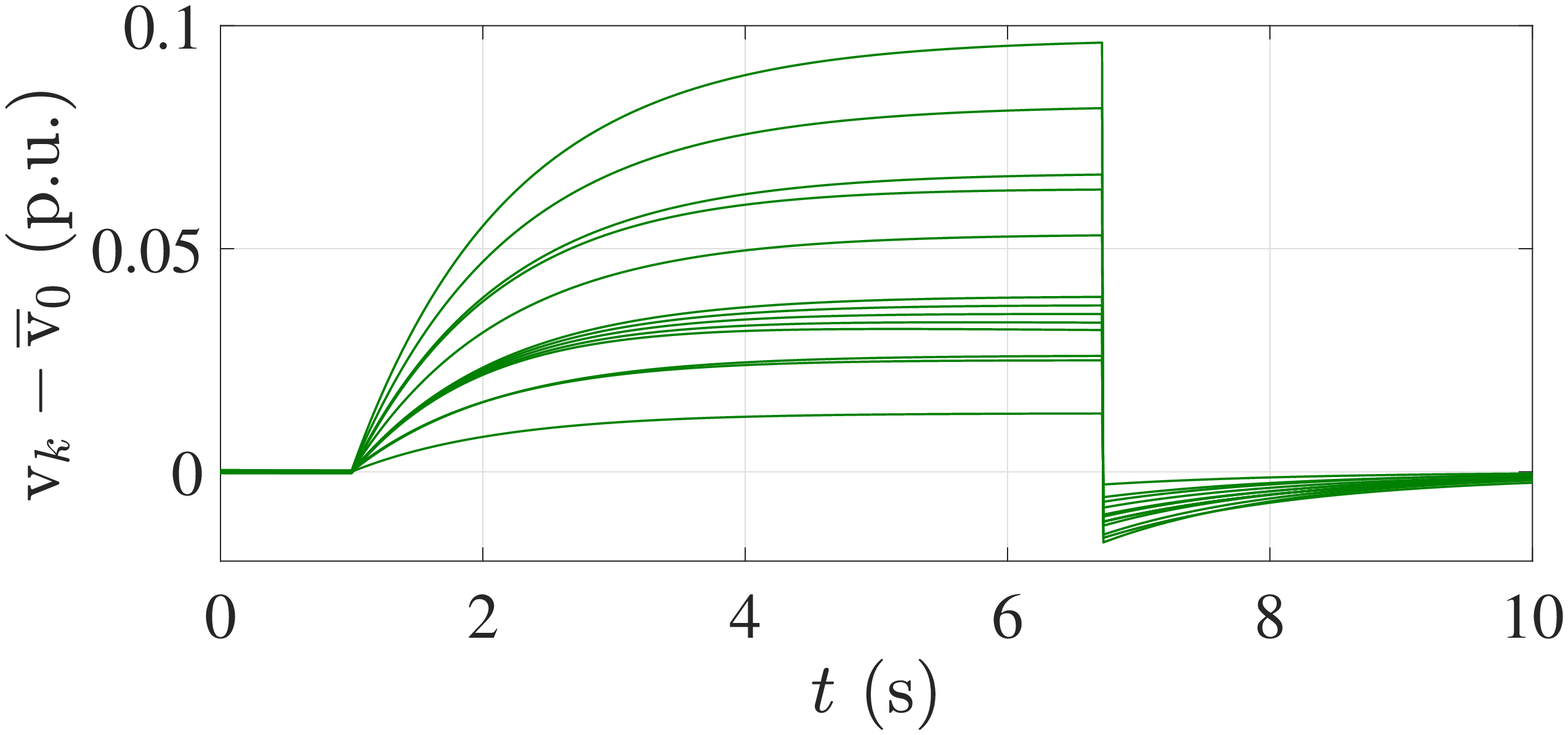}\label{subfig:vmd_naive}
}
\subfloat[][]{
\includegraphics[width=.33\linewidth]{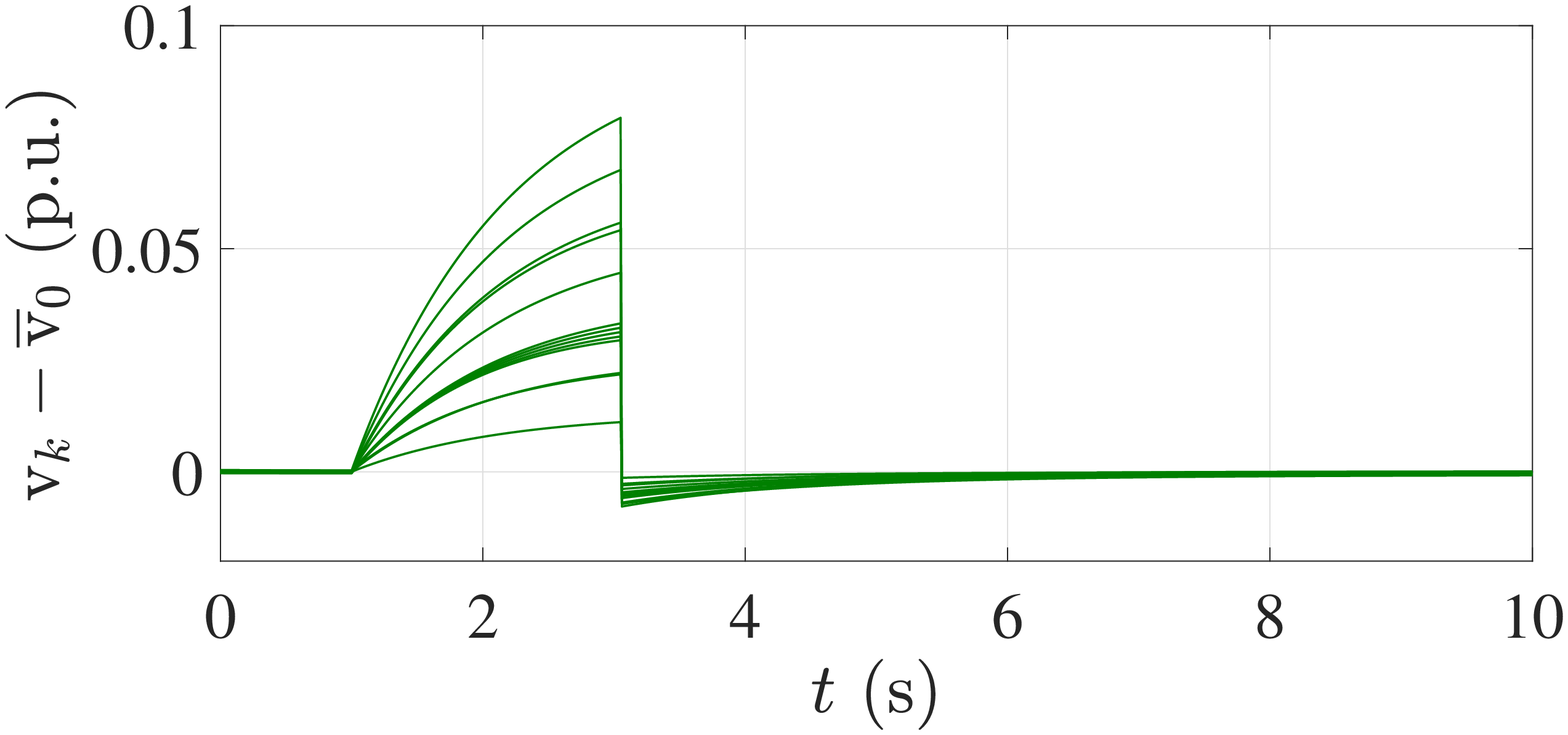}\label{subfig:vmd_low}
}
\subfloat[][]{
\includegraphics[width=.33\linewidth]{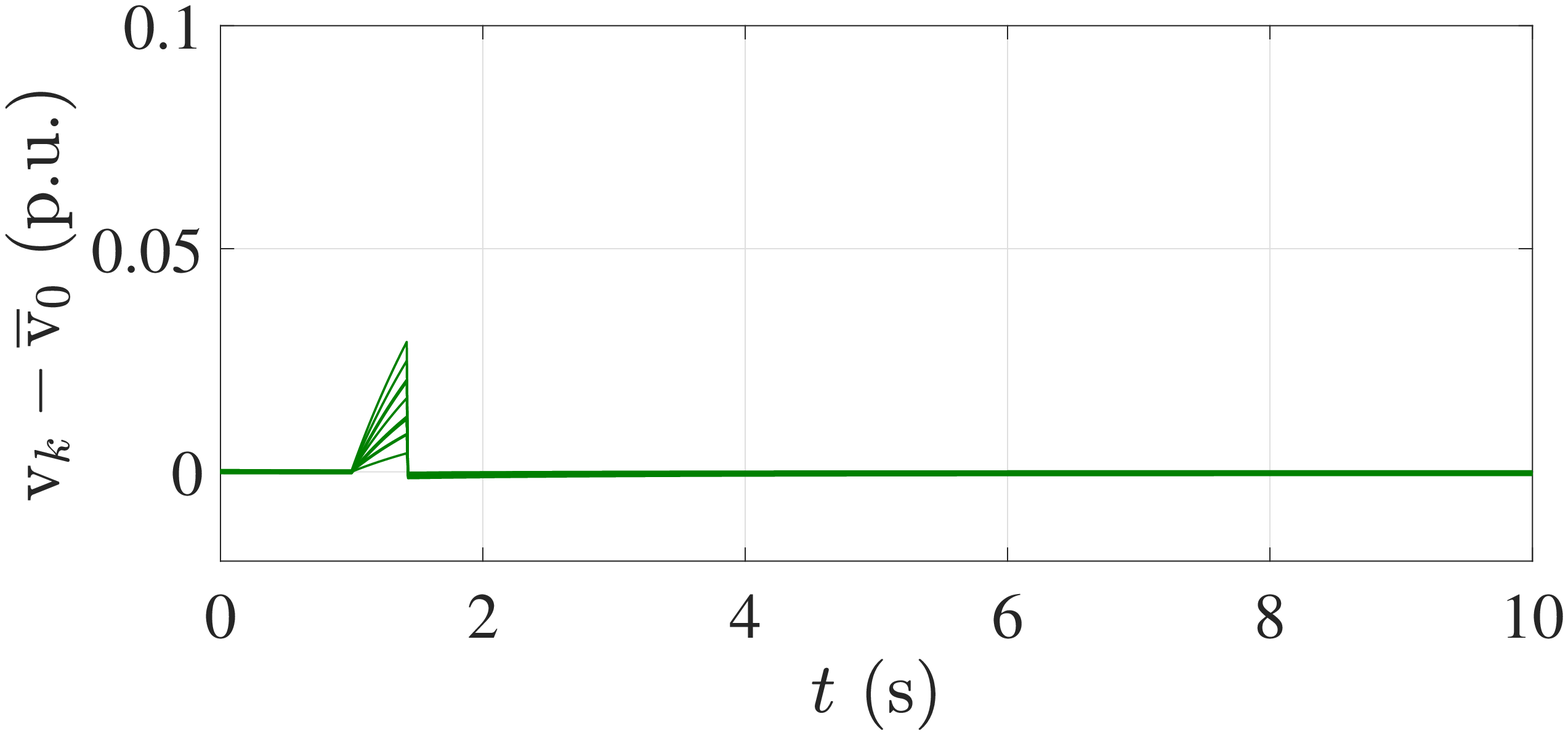}\label{subfig:vmd_high}
}
\caption{Time series of the voltage magnitude deviations ${\rm v}_k-{\rm v}_0$ in per unit at all buses.
(a): Naive approach.
(b): Proposed approach with the low gain.
(c): Proposed approach with the high gain.
}
\label{graph:vmd_scenario1}
\end{figure*}

\emph{Attack Detection:}
First, we investigate detection capability of the proposed residual generator under disconnection.
Suppose that the filter $S_i=I$ for any local residual generator.
The time series of the Euclidean norm of the residuals in per unit (p.u.) under measurement noise are illustrated in Fig.~\ref{graph:error_scenario1}, where the base unit is taken to be the detection threshold, i.e., $\gamma_i=1$.
The detection time instants with the naive approach, the proposed approach with the low gain, and that with the high gain are $6.51$~s, $2.98$~s, and $1.49$~s, respectively.
This figure verifies the theoretical finding that the stability of the residual generator is preserved under disconnection.
It can also be confirmed that the detection time is shorter as the feedback gain increases.
The time series of the voltage magnitudes of all buses in p.u. are illustrated in Fig.~\ref{graph:vmd_scenario1} where the base unit is taken to be the reference voltage.
It is indicated that early attack detection helps in suppression of voltage magnitude deviation caused by malicious actions.

\begin{figure*}[t]
\centering
\subfloat[][]{
\includegraphics[width=.33\linewidth]{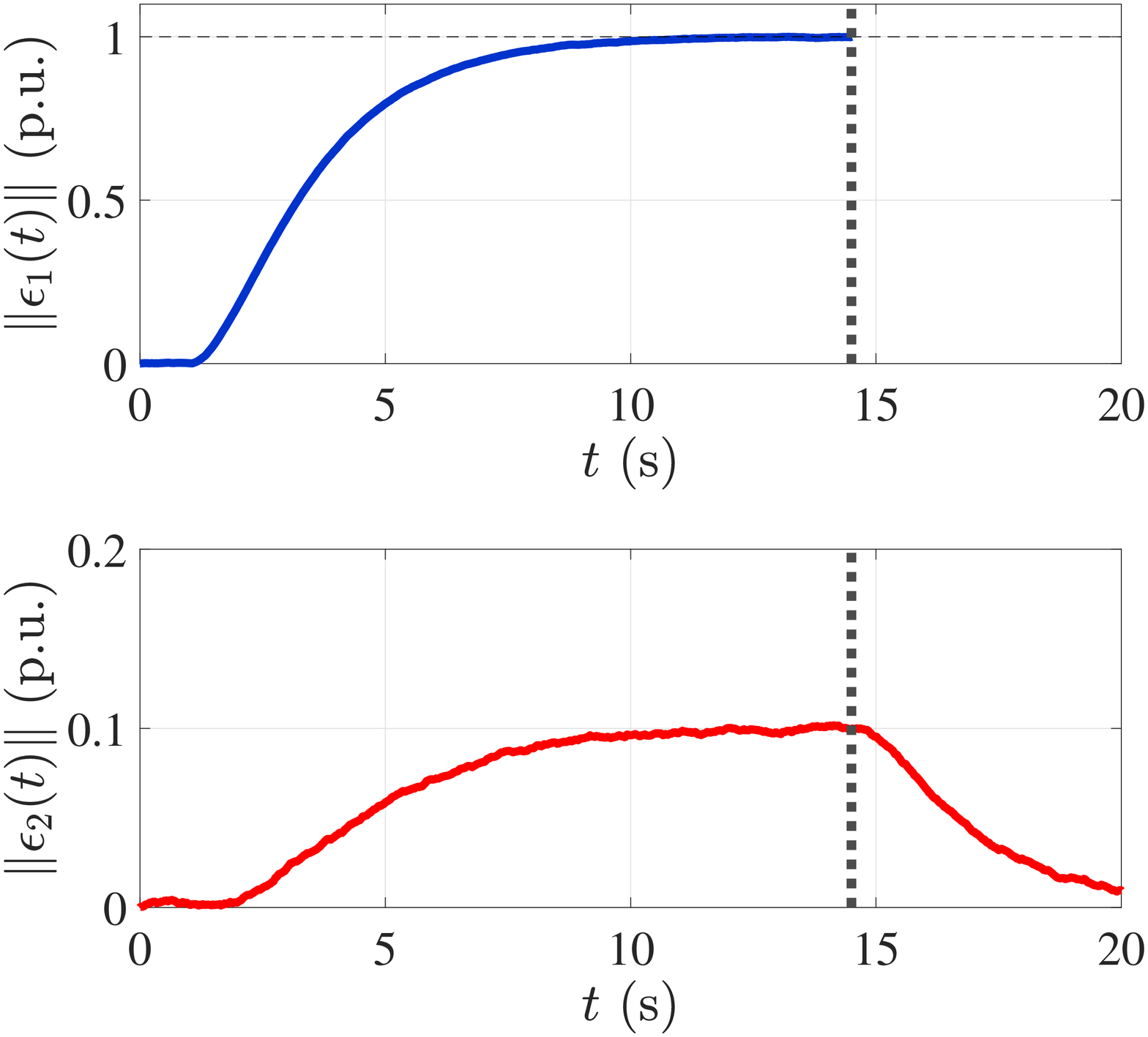}\label{subfig:err2_naive}
}
\subfloat[][]{
\includegraphics[width=.33\linewidth]{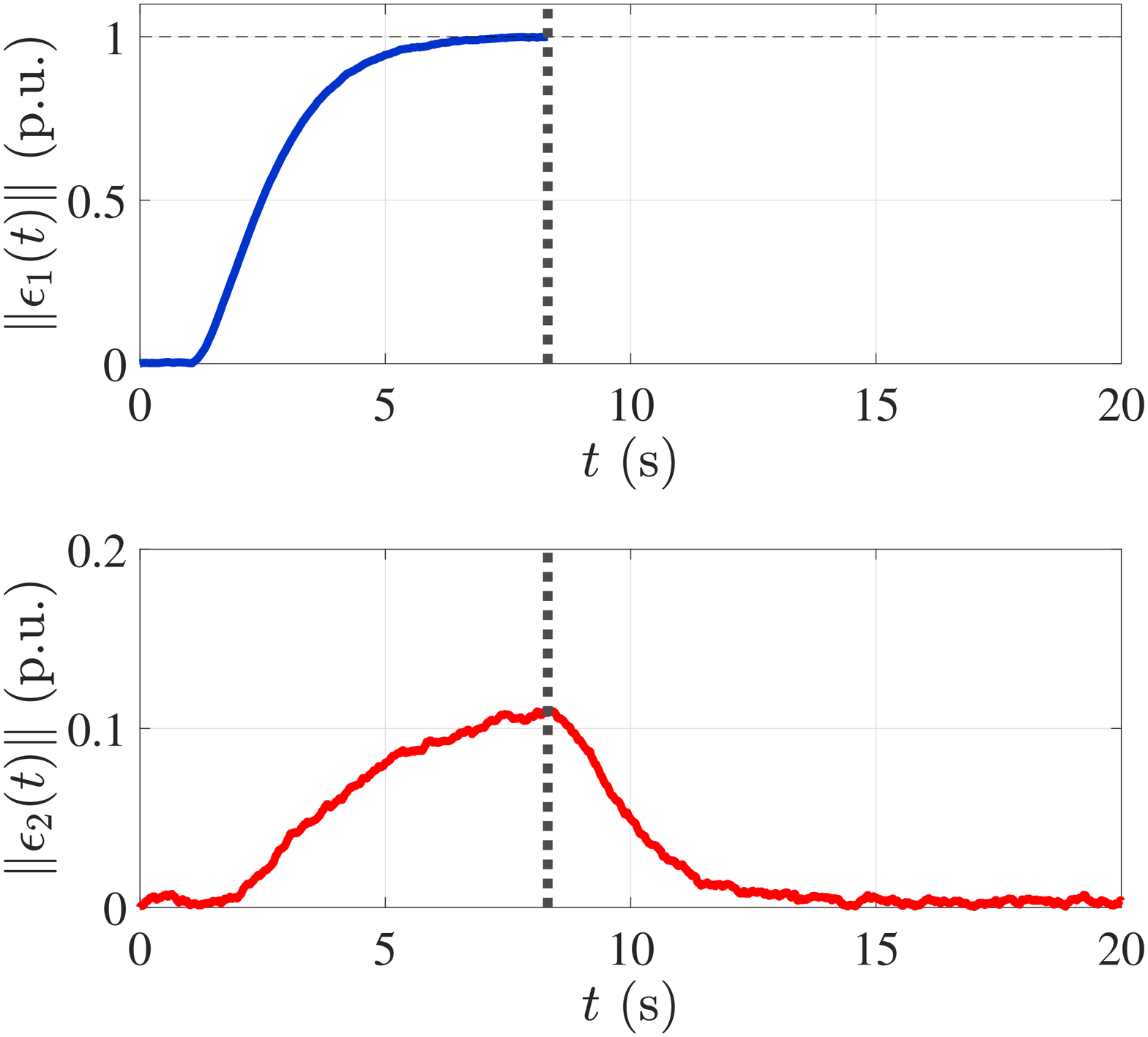}\label{subfig:err2_low}
}
\subfloat[][]{
\includegraphics[width=.33\linewidth]{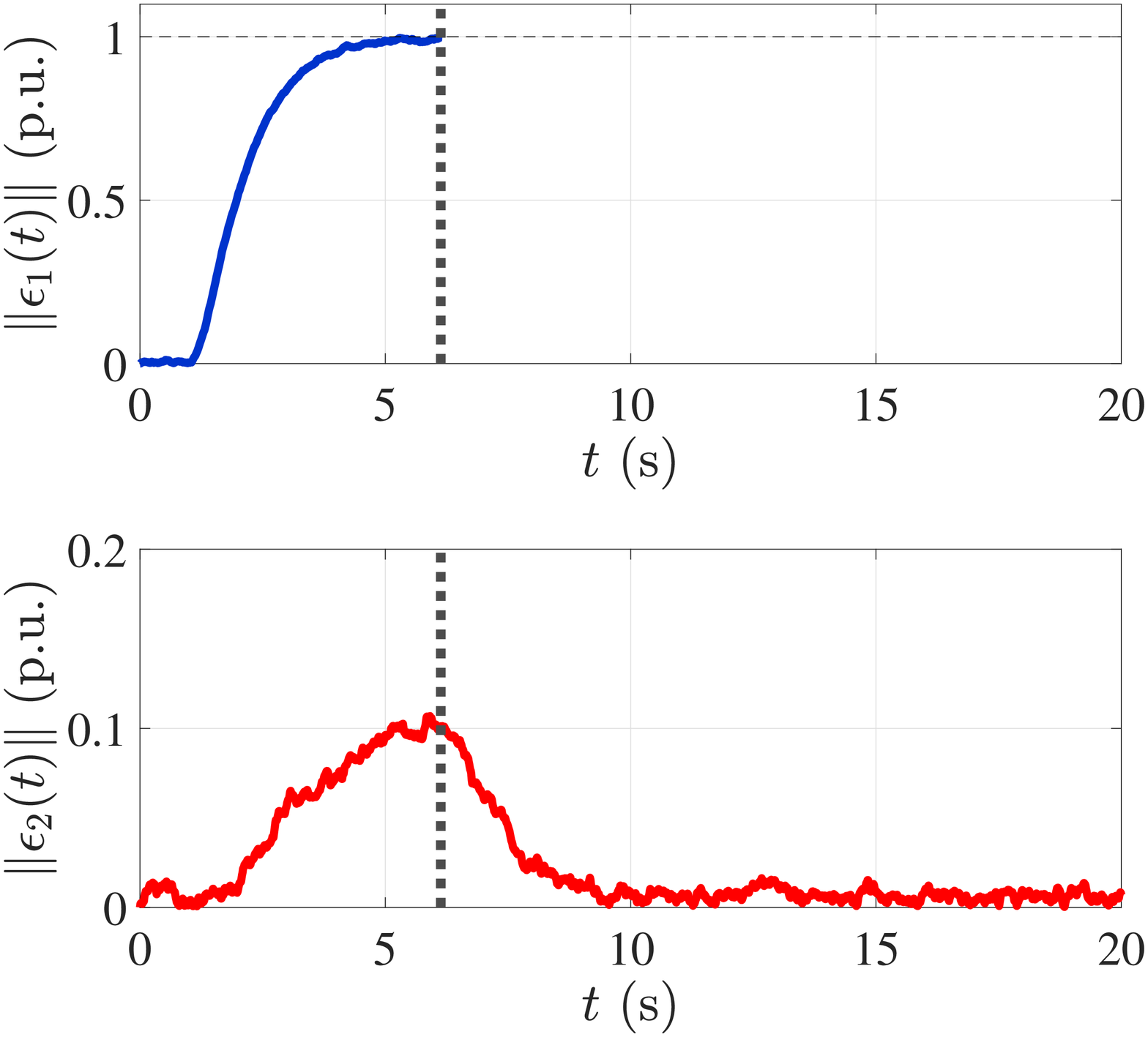}\label{subfig:err2_high}
}
\caption{Time series of $\|\epsilon_i(t)\|$ in per unit obtained with the residual generators with the noise reduction filter.
(a): Naive approach.
(b): Proposed approach with the low gain.
(c): Proposed approach with the high gain.
}
\label{graph:error_scenario2}
\end{figure*}

It can also be observed from the bottom subfigures of Fig.~\ref{graph:error_scenario1} that the effect of the noise is enlarged as the feedback gain increases.
To reduce noise effects, consider designing $S_i$ to be a noise reduction filter.
The $i$th filter is designed by $S_i={\rm diag}(\Psi_{\rm k})$ where $\Psi_k$ is the second-order Bessel filter with the cutoff frequency $1$~Hz, which is twice of the reciprocal of the inverter's time constant.
The time series of the Euclidean norm of the residuals with the noise reduction filter is depicted by Fig.~\ref{graph:error_scenario2}.
It can be observed that the noise is significantly reduced by the filter.
Furthermore, the detection time of the proposed residual generator is still faster than that of the naive approach.
It should be noted that the time constant of the residual generator is made large owing to the noise reduction filter.
As a result, the detection time is longer compared to the case of Fig.~\ref{graph:error_scenario1}.

\begin{figure*}[t]
\centering
\subfloat[][]{
\includegraphics[width=.33\linewidth]{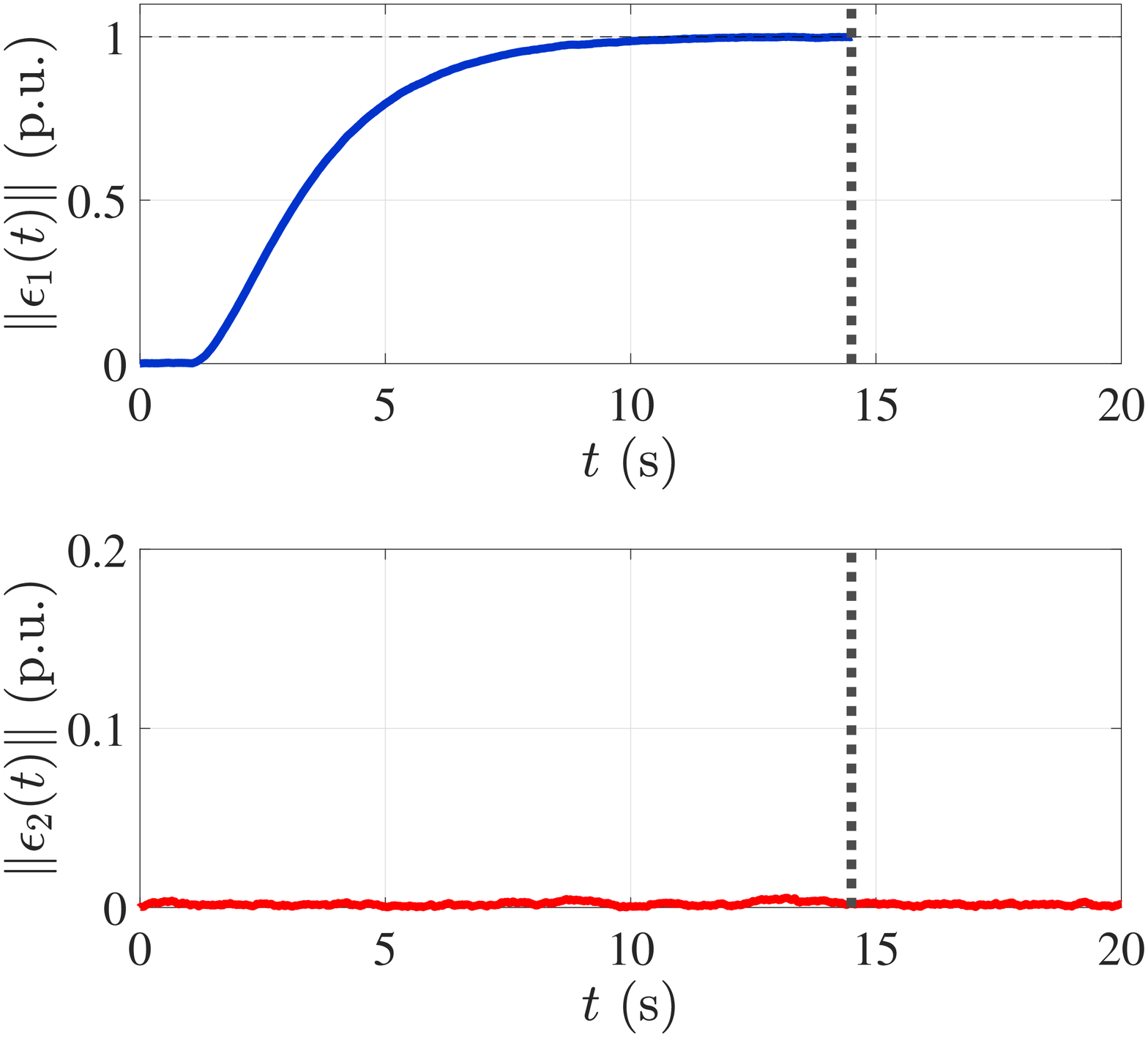}\label{subfig:err3_naive}
}
\subfloat[][]{
\includegraphics[width=.33\linewidth]{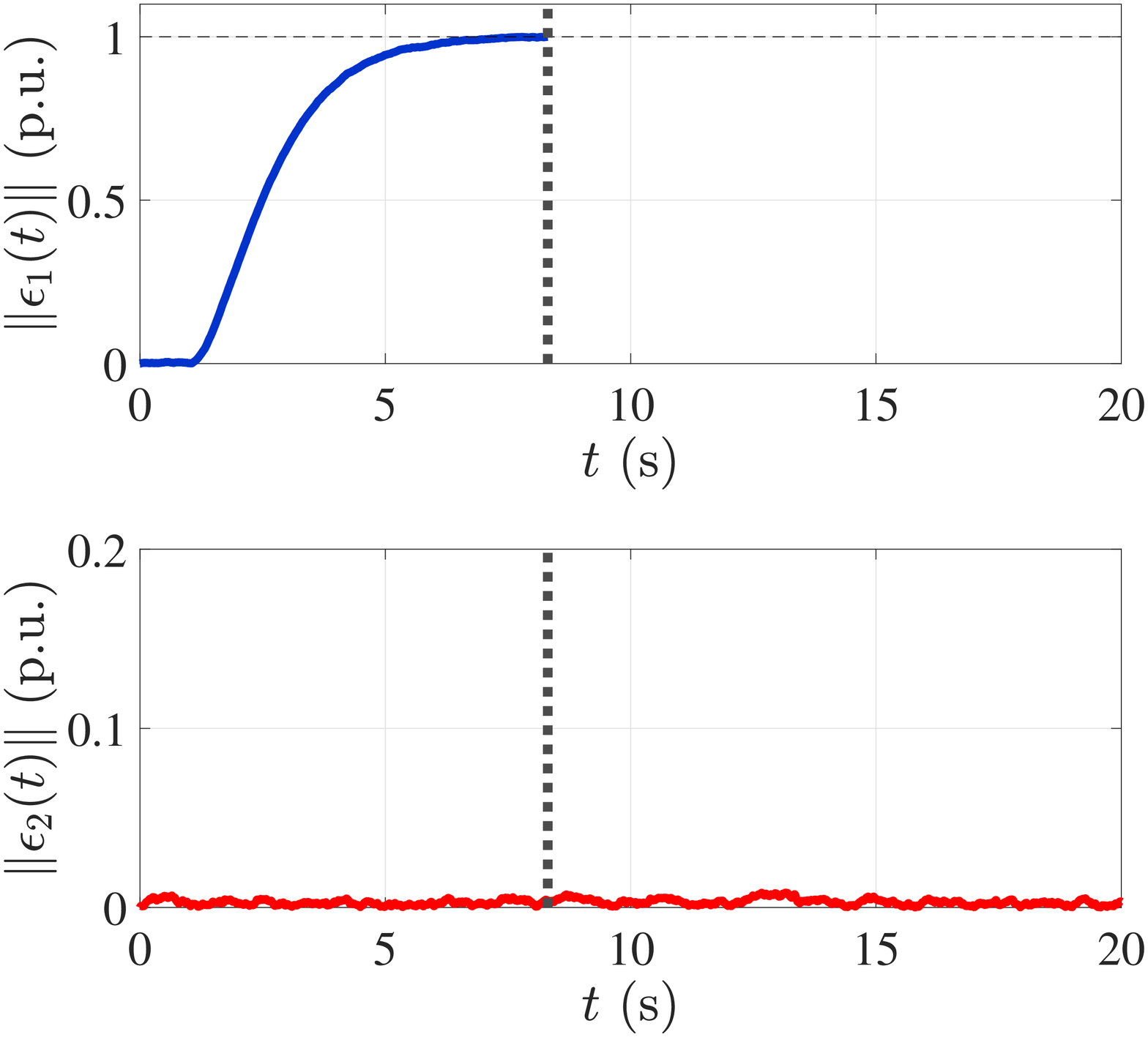}\label{subfig:err3_low}
}
\subfloat[][]{
\includegraphics[width=.33\linewidth]{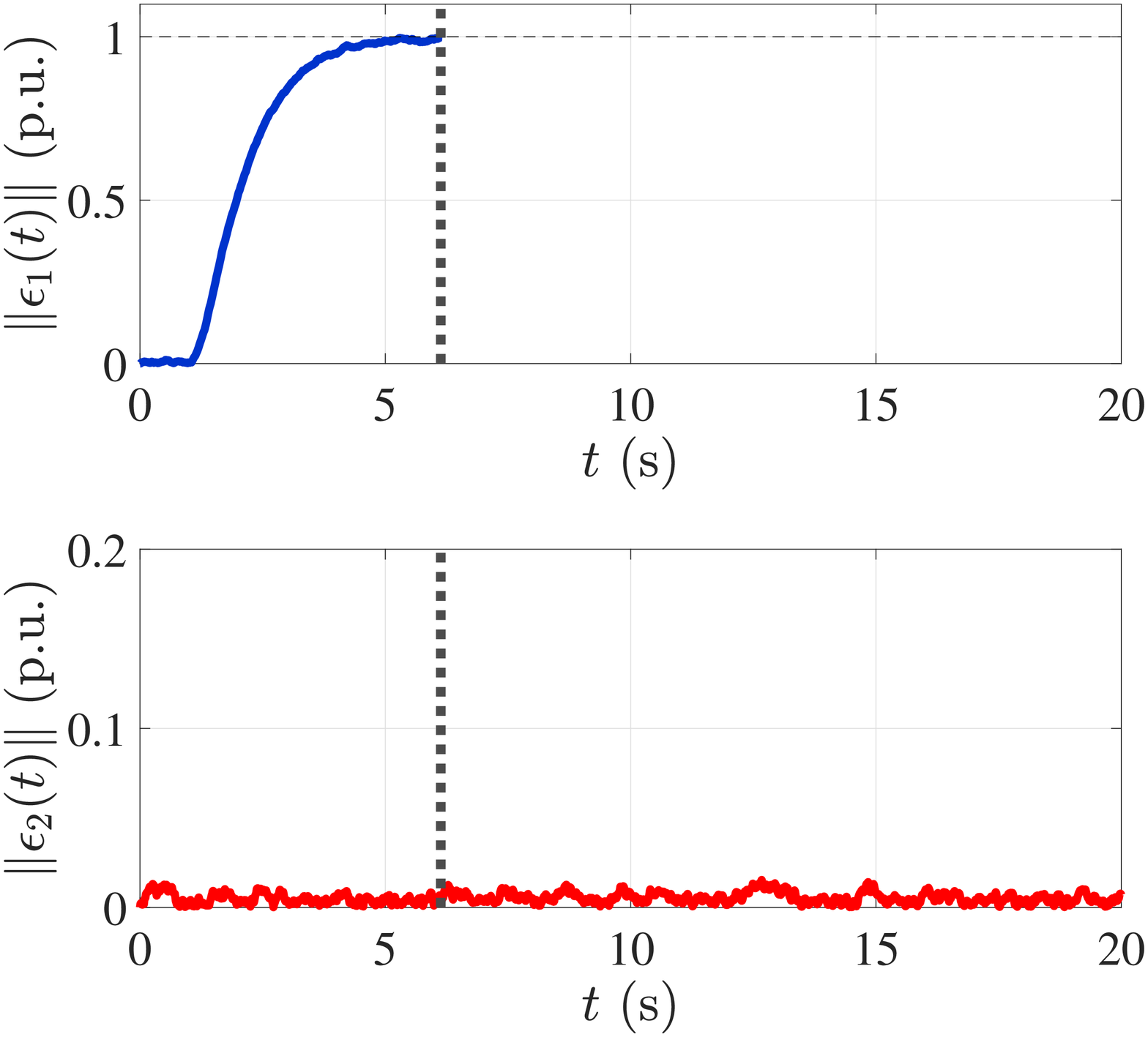}\label{subfig:err3_high}
}
\caption{Time series of $\|\epsilon_i(t)\|$ in per unit obtained with the residual generators with the filter composed of the isolation filter and the noise reduction filter.
(a): Naive approach.
(b): Proposed approach with the low gain.
(c): Proposed approach with the high gain.
}
\label{graph:error_scenario3}
\end{figure*}

\emph{Attack Isolation:}
It is found at the bottom subfigures of Fig.~\ref{graph:error_scenario2} that the residual signal in terms of the non-attacked subsystem is excited by the attack.
Although the amplitude is not very large, those residual signals can be a cause of misidentification of the attacked subsystem.
To enhance the identification capability, we design an isolation filter with noise reduction.
It can be confirmed that this partition satisfies the existence condition on an isolation filter in Theorem~\ref{thm:isolation}.
The unknown input observer, explained in Appendix~\ref{app:iso}, is employed for the isolation and the entire filter is constructed as the cascaded system with the isolation filter and the Bessel filter designed in the previous example.
The time series of the Euclidean norm of the residuals with the filter is depicted by Fig.~\ref{graph:error_scenario3}.
This figure indicates that the residual signals outside the attacked subsystem excited by the attack are almost completely removed by the isolation filter without delay of detection.
As claimed in the theoretical results, the attack isolation is successfully performed.

In summary, the numerical examples indicate that the proposed detector can preserve its detection and isolation capability under attack containment in an incident handling process.

\section{Conclusion}
\label{sec:conc}

Incident handling is a crucial notion for coping with adverse events.
This study has treated incident handling for networked control systems and pointed out the importance of the containment step,
which is carried out by disconnecting attacked components.
Network topology change caused by disconnection can lead to loss of detection capability of the equipped model-based attack detector.
Separation-based reconfiguration has been proposed and the disconnection-aware attack detection and isolation problems have been addressed.
A design method of a distributed residual generator based on retrofit control has been developed.
Its practical impacts are verified through numerical examples of low-voltage distribution networks.

An important direction for future work is attack detector design,
which considers performance of the detection unit, such as true and false alarm rates.
In addition, network reconfiguration after mitigation of the attack is another direction.
Finally, the procedure of disconnecting components has not been discussed in detail.
In practice, the unit that executes the disconnection is also a part of the networked control system,
and hence a security framework including the disconnection unit with combination of network security techniques is needed.

\appendices

\section{Detection without Knowledge of Initial State}
\label{app:detection_without_initial_state}

When the initial state is unknown, there exists a possibility for undetectable attacks that take advantage of this lack, called zero-dynamics attack~\cite{Pasqualetti2015Control}.
Zero-dynamics attacks are carried out by injecting a particular signal that excites the zero dynamics of the system to be protected.
When the zero dynamics is stable, zero-dynamics attacks are generally not very threatening because those effects are diminishing by themselves.
However, if the system has unstable zeros, the state can diverge by zero-dynamics attacks without being detected.
Such unstable zeros should be eliminated through modification of system architecture.
This appendix shows that our proposed residual generator does not create additional unstable zeros.

The definition of invariant zeros, which is a standard notion of zeros in the state-space representation, is as follows~\cite{Schrader1989Research}.
\begin{defin}[Invariant Zero]
Consider a state-space representation of a linear time-invariant system with the matrices $(A,B,C,D)$.
The invariant zeros of the system are defined to be the complex numbers $s_0$ such that
\[
 {\rm rank}\left[
 \begin{array}{cc}
 A-s_0I & B\\
 C & D
 \end{array}
 \right]<n+{\rm min}(m,p)
\]
where $n,m,p$ denote the dimensions of the state, input, and output, respectively.
When the real part of $s_0$ is negative and nonnegative, $s_0$ is called a stable and unstable zero, respectively.
\end{defin}



Observe that
\[
\epsilon_\mathcal{I} ={\rm diag}(S_iM_i)\left({\rm diag}(G_{y_i v_i}) v_\mathcal{I}+{\rm diag}(G_{y_i a_i})a_\mathcal{I}\right)
\]
where the $i$th block diagonal component $M_i$ represents an observer for the $i$th subsystem.
Because an observer with static error feedback does not move zeros, we expect that the entire system has no unstable zeros if the $i$th subsystem and the system from $a_\mathcal{I}$ to $v_\mathcal{I}$ have no unstable zeros.
Let $T_{v_\mathcal{I} a_\mathcal{I}}$ denote the system from $a_\mathcal{I}$ to $v_\mathcal{I}$.
The following assumption is made.
\begin{assum}\label{assum:stable_zeros}
The system $T_{v_\mathcal{I} a_\mathcal{I}}$, whose state-space form is consistent with~\eqref{eq:ori_sys} and~\eqref{eq:ori_interaction}, has no unstable zeros for any $\mathcal{I} \in \mathfrak{I}$.
\end{assum}

Under Assumption~\ref{assum:stable_zeros}, the following theorem holds.
\begin{theorem}\label{thm:stable_zeros}
Let Assumptions~\ref{assum:ori_sta} and~\ref{assum:stable_zeros} hold.
Assume that all subsystems are stable.
Consider the residual generator in Theorem~\ref{thm:detection}.
If $S_{\mathcal{I}}$ has no unstable zeros,
then all invariant zeros of the system from $a_\mathcal{I}$ to $\epsilon_\mathcal{I}$ are stable for any $\mathcal{I} \in \mathfrak{I}$.
\end{theorem}
\begin{IEEEproof}
Observe that the system from $a_\mathcal{I}$ to $\epsilon_\mathcal{I}$ can be represented by $S_\mathcal{I}M_\mathcal{I}T_{y_\mathcal{I} a_\mathcal{I}}$.
It suffices to show that
\[
\begin{array}{l}
 S_\mathcal{I}M_\mathcal{I}T_{y_\mathcal{I} a_\mathcal{I}}\\
 = S_\mathcal{I}M_\mathcal{I}\left[{\rm diag}(G_{y_i v_i})\ {\rm diag}(G_{y_i a_i})\right]
 \left[
 \begin{array}{c}
 T_{v_{\mathcal{I}} a_{\mathcal{I}}}\\
 I
 \end{array} 
 \right]
\end{array}
\]
has no invariant zeros in the time domain.
Because a cascaded system composed of two systems that have no unstable zeros does not have unstable zeros, from the assumptions, it suffices to show that $M_\mathcal{I}{\rm diag} (G_{y_i (v_i, a_i)})$ has no unstable zeros.
Owing to its block diagonal structure, we show that $M_iG_{y_i (v_i,a_i)}$ has no unstable zeros.

Noticing that invariant zeros are invariant under coordinate transformation,
we take $e_i:=x_i-\hat{x}_i$
and $f_i:=x_i+\hat{x}_i$ with $\hat{\chi}_i$ itself.
Then the realization of $M_iG_{y_i (v_i,a_i)}$ with this coordination can be described by
\[
 \left\{
 \begin{array}{l}
 \dot{e}_i  = (A_i-H_iC_i)e_i + (U'_i-H_iV'_i)v'_i\\
 \dot{f}_i  = A_i f_i + H_iC_i e_i + (U'_i+H_iV'_i)v'_i\\
 \dot{\hat{\chi}} = A_i\hat{\chi} + H_iC_ie_i + H_iV'_i v'_i\\
 y-\hat{y}_i = C_ie_i + V'_i v'_i
 \end{array}
 \right.
\]
where $U'_i:=[U_i\ X_i], V'_i:=[V_i\ Y_i],$ and $v'_i$ is the stacked vector of $v_i$ and $a_i$.
Then we have
\[
\arraycolsep=4pt
\begin{array}{l}
 \left[
 \begin{array}{cccc}
 A_i-H_iC_i-sI & 0 & 0 & U'_i-H_iV'_i\\
 H_iC_i & A_i-sI & 0 & U'_i+H_iV'_i\\
 H_iC_i & 0 & A_i-sI & H_iV'_i\\
 C_i & 0 & 0 & V'_i
 \end{array}
 \right]\\
 =
 \left[
 \begin{array}{cccc}
 I & 0 & 0 & -H_i\\
 0 & I & 0 & H_i\\
 0 & 0 & I & H_i\\
 0 & 0 & 0 & I
 \end{array}
 \right]
 \left[
 \begin{array}{cccc}
 A_i-sI & 0 & 0 & U'_i\\
 H_iC_i & A_i-sI & 0 & U'_i\\
 H_iC_i & 0 & A_i-sI & 0\\
 C_i & 0 & 0 & V'_i
 \end{array}
 \right].
\end{array}
\]
From Assumption~\ref{assum:stable_zeros}, the rank of this matrix is deficient if and only if $s$ is an eigenvalue of $A_i$.
From the assumption, all eigenvalues of $A_i$ are stable.
Thus the invariant zeros are stable, which proves the claim.
\end{IEEEproof}

Theorem~\ref{thm:stable_zeros} shows the validity of our approach even without knowledge of initial state.

\section{Brief Review of Retrofit Control}
\label{app:retro}

\begin{figure}[t]
\centering
\subfloat[][]{\includegraphics[width=.48\linewidth]{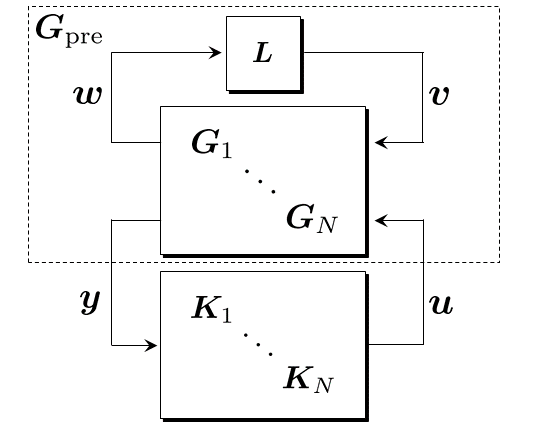}\label{subfig:sys_retro}} \quad
\centering
\subfloat[][]{\includegraphics[width=.48\linewidth]{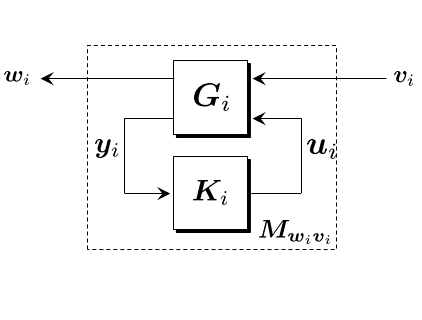}\label{subfig:M_i}}\\
\subfloat[][]{\includegraphics[width=.96\linewidth]{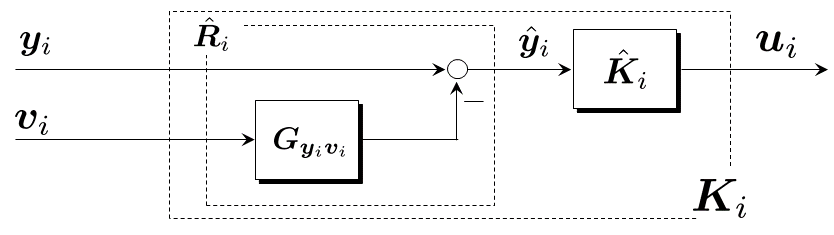}\label{subfig:retro_str}}
\caption[]{Block diagrams relevant to retrofit control.
(a): Block diagram of the networked system of interest in the retrofit control framework.
(b): Block diagram from the viewpoint of the $i$th subcontroller designer.
(c): Internal structure of all output-rectifying retrofit controllers.
}
\label{fig:sys_retro}
\end{figure}

Retrofit control has originally been proposed for facilitating modular design of a control system, namely, independent design of subcontrollers only with its corresponding subsystem model in a networked system.
The networked system of interest in the retrofit control framework is depicted in Fig.~\ref{subfig:sys_retro},
where $\boldsymbol{G}_1,\ldots,\boldsymbol{G}_N$ denote the transfer matrices of subsystems, $\boldsymbol{L}$ denotes the transfer matrix of the interaction, and $\boldsymbol{K}_1,\ldots,\boldsymbol{K}_N$ denote subcontrollers to be designed.
In the retrofit control framework, it is supposed that there are $N$ subcontroller designers each of whom is responsible for designing her corresponding subcontroller $\boldsymbol{K}_i$ only with the model information on her own subsystem $\boldsymbol{G}_i$.
A crucial premise of retrofit control is that the preexisting system $\boldsymbol{G}_{\rm pre}$ is internally stable, where $\boldsymbol{G}_{\rm pre}$ represents the interconnected system composed only of ${\rm diag}(\boldsymbol{G}_i)$ and the interaction $\boldsymbol{L}$ without the controller ${\rm diag}(\boldsymbol{K}_i)$.
The objective of the subcontroller designers is to improve a desirable control performance while preserving the stability.

A block diagram of the isolated feedback system from the viewpoint of the $i$th subcontroller designer is depicted in Fig.~\ref{subfig:M_i}.
The transfer matrix $\boldsymbol{M}_{\boldsymbol{w}_i\boldsymbol{v}_i}$ denotes the input-output map from $\boldsymbol{v}_i$ to $\boldsymbol{w}_i$ with the feedback controller $\boldsymbol{K}_i$.
The difficulty for designing $\boldsymbol{K}_i$ in Fig.~\ref{subfig:M_i} is that stability of the entire networked system in Fig.~\ref{subfig:sys_retro} may be lost even if $\boldsymbol{K}_i$ stabilizes the local closed-loop system.
The fundamental idea of retrofit control is to design each subcontroller so as to maintain the closed-loop relationship between its corresponding interaction signals to be invariant.
The mathematical description of this idea is given by
\begin{equation}\label{eq:Mi}
 \boldsymbol{M}_{\boldsymbol{w}_i\boldsymbol{v}_i}=\boldsymbol{G}_{\boldsymbol{w}_i\boldsymbol{v}_i}
\end{equation}
where $\boldsymbol{G}_{\boldsymbol{w}_i\boldsymbol{v}_i}$ denotes the submatrix of $\boldsymbol{G}_i$ with respect to $\boldsymbol{w}_i$ and $\boldsymbol{v}_i$.

We refer to the controllers that stabilize $\boldsymbol{G}_i$ and satisfy~\eqref{eq:Mi} as \emph{retrofit controllers.}
By means of retrofit controllers, modular design of subcontrollers can be achieved as shown in the following proposition~\cite{Ishizaki2019Modularity}.
\begin{prop}\label{prop:retro}
Assume that $\boldsymbol{G}_{\rm pre}$ is internally stable and $\boldsymbol{G}_i$ is stable for $i=1,\ldots,N$.
If $\boldsymbol{K}_i$ stabilizes $\boldsymbol{G}_i$ and satisfies~\eqref{eq:Mi} for any $i=1,\ldots,N$,
then the networked control system in Fig.~\ref{subfig:sys_retro} is internally stable.
\end{prop}

Note that the conditions become also necessary when the model of the other subsystems and the interaction are completely unknown to the subcontroller designer~\cite{Ishizaki2019Modularity}.
Note also that the same idea can be adopted even without the technical assumption on stability of every subsystem $\boldsymbol{G}_i$ although~\eqref{eq:Mi} takes a more cumbersome form~\cite{Ishizaki2019Modularity}.

The condition in Proposition~\ref{prop:retro} can equivalently be rewritten by
\begin{equation}\label{eq:GQG}
 \boldsymbol{G}_{\boldsymbol{w}_i\boldsymbol{u}_i}\boldsymbol{Q}_i\boldsymbol{G}_{\boldsymbol{y}_i\boldsymbol{v}_i}=0
\end{equation}
where $\boldsymbol{Q}_i:=(I-\boldsymbol{K}_i\boldsymbol{G}_{\boldsymbol{y}_i\boldsymbol{u}_i})^{-1}\boldsymbol{K}_i \in \mathcal{RH}_{\infty}$ is the Youla parameter of $\boldsymbol{K}_i$ and $\boldsymbol{G}_{\boldsymbol{w}_i\boldsymbol{u}_i},\boldsymbol{G}_{\boldsymbol{y}_i\boldsymbol{v}_i},\boldsymbol{G}_{\boldsymbol{y}_i\boldsymbol{u}_i}$ are the submatrices of $\boldsymbol{G}_i$ with respect to the subscript signals.
A sufficient condition for~\eqref{eq:GQG} given by
\begin{equation}\label{eq:KG}
 \boldsymbol{Q}_i \boldsymbol{G}_{\boldsymbol{y}_i\boldsymbol{v}_i}=0
\end{equation}
provides a particular class of retrofit controllers, referred to as \emph{output-rectifying retrofit controllers}, which are defined as the controllers such that the corresponding Youla parameter $\boldsymbol{Q}_i\in \mathcal{RH}_{\infty}$ satisfies~\eqref{eq:KG}.
This class of retrofit controllers is tractable in the sense that all output-rectifying retrofit controllers can explicitly be parameterized with a free parameter when the interaction signal $\boldsymbol{v}_i$ is measurable as shown in the following proposition~\cite{Ishizaki2019Modularity}.
\begin{prop}\label{prop:out_str}
Assume that the interaction signal $\boldsymbol{v}_i$ is measurable in addition to the measurement output $\boldsymbol{y}_i$.
Then $\boldsymbol{K}_i$ is an output-rectifying retrofit controller if and only if there exists an internal controller $\hat{\boldsymbol{K}}_i$ such that
\[
 \boldsymbol{K}_i=\hat{\boldsymbol{K}}_i\hat{\boldsymbol{R}}_i,\quad \hat{\boldsymbol{Q}}_i:=(I-\hat{\boldsymbol{K}}_i\boldsymbol{G}_{\boldsymbol{y}_i\boldsymbol{u}_i})^{-1}\hat{\boldsymbol{K}}_i \in \mathcal{RH}_{\infty}
\]
with $\hat{\boldsymbol{R}}_i=[I\  {-\boldsymbol{G}_{\boldsymbol{y}_i\boldsymbol{v}_i}}]$.
\end{prop}

Proposition~\ref{prop:out_str} implies that an internal structure of all output-rectifying retrofit controllers is illustrated by Fig.~\ref{subfig:retro_str}.
One of the features observed in this structure is that the control input $\boldsymbol{u}_i$ is generated through a locally stabilizing controller $\hat{\boldsymbol{K}}_i$ with a ``rectified'' measurement $\hat{\boldsymbol{y}}_i$, which is given by
\[
 \hat{\boldsymbol{y}}_i=\hat{\boldsymbol{R}}_i[\boldsymbol{y}_i^{\sf T}\ \boldsymbol{v}_i^{\sf T}]^{\sf T}=\boldsymbol{y}_i-\boldsymbol{G}_{\boldsymbol{y}_i\boldsymbol{v}_i}\boldsymbol{v}_i,
\]
where the effects of $\boldsymbol{v}_i$ to $\boldsymbol{y}_i$ are eliminated in $\hat{\boldsymbol{R}}_i$.
This rectification is the essential technique for constructing an output-rectifying retrofit controller.
The assumption on interaction measurability can be fulfilled by introducing additional sensors in practical applications.

On the other hand, we can consider another class of retrofit controllers that have a dual form of~\eqref{eq:KG}.
Define \emph{input-rectifying retrofit controllers} as the controllers whose Youla parameter $\boldsymbol{Q}_i\in \mathcal{RH}_{\infty}$ satisfies
\begin{equation}\label{eq:GQ_gen}
 \boldsymbol{G}_{\boldsymbol{w}_i\boldsymbol{u}_i}\boldsymbol{Q}_i=0.
\end{equation}
However, compared to~\eqref{eq:KG}, this condition is difficult to exploit for control of physical systems.
Observe that~\eqref{eq:GQ_gen} has only the trivial solution $\boldsymbol{Q}_i=0$ when  the dimension of the input signal $\boldsymbol{u}_i$ is less than that of $\boldsymbol{w}_i$ for a full-column rank transfer matrix $\boldsymbol{G}_{\boldsymbol{w}_i\boldsymbol{u}_i}$.
Although we have to create additional input ports, e.g., introducing new actuators, to increase the dimension of $\boldsymbol{u}_i$,
equipment of actuators requires more efforts than that of sensors in general.
For the reason, no specific design methods for input-rectifying retrofit controllers have been developed.

Indeed, the feedback architecture inside the proposed distributed observer satisfies~\eqref{eq:GQ_gen}.
An important observation is that the control signals inside our distributed observer are not physical signals but \emph{cyber} signals,
and hence we can freely adjust its parameters such as corresponding injection ports as conducted in our proposed method.
This beneficial property enables us to apply the input-rectifying retrofit controller structure to the problem addressed in this study.

\section{Demonstration of Isolation Filter Design}
\label{app:iso}

This appendix demonstrates a design procedure of an isolation filter using unknown input observers (UIOs).
Although its existence condition is slightly stricter than the general isolation filter, an intuitive design algorithm is available and their internal structure is easy to understand.
Our objective is to design $S_i\in\mathcal{RH}_{\infty}$ such that
\[
 S_iM_iG_{y_i a_i} {\rm\ is\ left\ invertible},\ {\rm and}\ S_iM_iG_{y_i v_i}=0
\]
for $i=1,\ldots,N$.

An isolation filter using a UIO is designed as follows:
Fix $i\in\{1,\ldots,N\}$.
Let
\[
 \dot{z}_i=\tl{A}_iz_i+\tl{U}_iv_i,\quad \tl{y}_i=\tl{C}_iz_i
\]
be a realization of $M_iG_{y_iv_i}$.
Without loss of generality, the input matrix $\tl{U}_i$ is assumed to be full-column rank.
Consider
\begin{equation}\label{eq:UIO}
 \dot{\zeta}_i=\tl{F}_i\zeta_i+\tl{K}_i\tl{y}_i,\quad \hat{z}_i=\zeta_i+\tl{H}_i\tl{y}_i,
\end{equation}
which is called a UIO when $z_i(t)-\hat{z}_i(t)\to 0$ as $t\to \infty$ for any $v_i$ under any initial condition.
Now we make the following assumptions:
\begin{description}
\item[{\bf A1.}] The matrix $\tl{C}_i\tl{U}_i$ is left invertible.
\item[{\bf A2.}] The pair $(\tl{F}_i,\tl{C}_i)$ is detectable, where $\tl{F_i}$ is given below.
\end{description}
Those assumptions are a necessary and sufficient condition for the existence of a UIO~\cite[Chapter~3]{Chen1999Robust}.
Let
\[
 \begin{array}{l}
 \tl{H}_i=\tl{U}_i((\tl{C}_i\tl{U}_i)^{\sf T}\tl{C}_i\tl{U}_i)^{-1}(\tl{C}_i\tl{U}_i)^{\sf T},\\
 \tl{F}_i=\tl{A}_i-\tl{H}_i\tl{C}_i\tl{A}_i,\quad \tl{K}_i=\tl{F}_i\tl{H}_i
 \end{array}
\]
and then it turns out that~\eqref{eq:UIO} is a UIO when $\tl{F}_i$ is stable.
Note that, it suffices to introduce estimation error feedback when $\tl{F}_i$ is unstable.

We consider designing an isolation filter $S_i$ using the UIO.
Because the UIO tracks the effect of $v_i$ to $z_i$ without knowledge of $v_i$, we have
\[
 (I-\tl{C}_iG_{{\rm UIO},i})M_iG_{y_iv_i}v_i=0
\]
for any $v_i$ where $G_{{\rm UIO},i}$ is the frequency-domain representation of~\eqref{eq:UIO}.
Hence, we have
\[
\tl{y}_i - \tl{C}_iG_{{\rm UIO},i}\tl{y}_i = (I-\tl{C}_iG_{{\rm UIO},i})M_iG_{y_ia_i}a_i.
\]
Now it turns out that $(I-\tl{C}_iG_{{\rm UIO},i})M_iG_{y_ia_i}$ is left invertible if the condition derived by Theorem~\ref{thm:isolation} holds, and hence
\[
 S_i=I-\tl{C}_iG_{{\rm UIO},i}
\]
can be taken as the desired isolation filter.

\ifCLASSOPTIONcaptionsoff
  \newpage
\fi

\bibliography{sshrrefs}
\bibliographystyle{IEEEtran}

\begin{IEEEbiography}{Hampei Sasahara}(M’15)
received the Ph.D. degree in engineering from Tokyo Institute of Technology in 2019.
He is currently a postdoctral researcher with KTH Royal Institute of Technology.
His main interest is secure cyber-physical-human system design.
\end{IEEEbiography}

\begin{IEEEbiography}{Takayuki Ishizaki}(M’10)
was born in Aichi, Japan, in 1985. He received the B.Sc., M.Sc., and Ph.D. degrees in Engineering from Tokyo Institute of Technology, Tokyo, Japan, in 2008, 2009, and 2012, respectively.
He served as a Research Fellow of the Japan Society for the Promotion of Science from April 2011 to October 2012.
From October to November 2011, he was a Visiting Student at Laboratoire Jean Kuntzmann, Universite Joseph Fourier, Grenoble, France. From June to October 2012, he was a Visiting Researcher at School of Electrical Engineering, Royal Institute of Technology, Stockholm, Sweden.
Since November 2012, he has been with Tokyo Institute of Technology, where he is currently an Associate Professor at the Department of Systems and Control Engineering.
His research interests include the development of network model reduction and its applications, retrofit control and its applications, and power systems control with distributed energy resources.
Dr. Ishizaki is a member of IEEE, SICE, and ISCIE. He was the recipient of several awards including Best Paper Awards from SICE in 2010 and from ISCIE in 2015, Finalist of the 51st IEEE CDC Best Student-Paper Award, and Pioneer Award of Control Division from SICE in 2019.
\end{IEEEbiography}

\begin{IEEEbiography}{Jun-ichi Imura}(SM’18)
received the M.E. degree in applied systems science and the Ph.D. degree in mechanical engineering from Kyoto University, Kyoto, Japan, in 1990 and 1995, respectively.
He served as a Research Associate at the Department of Mechanical Engineering, Kyoto University, from 1992 to 1996, and as an Associate Professor in the Division of Machine Design Engineering, Faculty of Engineering, Hiroshima University, Hiroshima, Japan, from 1996 to 2001.
From May 1998 to April 1999, he was a Visiting Researcher at the Faculty of Mathematical Sciences, University of Twente, Enschede, The Netherlands.
Since 2001, he has been with Tokyo Institute of Technology, Tokyo, Japan, where he is currently a Professor at the Department of Systems and Control Engineering.
His research interests include modeling, analysis, and synthesis of nonlinear systems, hybrid systems, and large-scale network systems with applications to power systems, ITS, biological systems, and industrial process systems.
He is a member of the Society of Instrument and Control Engineers (SICE), The Institute of Systems, Control and Information Engineers (ISCIE), and The Robotics Society of Japan.
\end{IEEEbiography}

\begin{IEEEbiography}{Henrik Sandberg}(SM'21)
received the M.Sc. degree in engineering physics and the Ph.D. degree in automatic control from Lund University, Lund, Sweden, in 1999 and 2004, respectively.
He is a Professor with the Division of Decision and Control Systems, KTH Royal Institute of Technology, Stockholm, Sweden.
From 2005 to 2007, he was a Postdoctoral Scholar with the California Institute of Technology, Pasadena, CA, USA.
In 2013, he was a Visiting Scholar with the Laboratory for Information and Decision Systems, Massachusetts Institute of Technology, Cambridge, MA, USA.
He has also held visiting appointments with the Australian National University, Canberra, ACT, USA, and the University of Melbourne, Parkville, VIC, Australia.
His current research interests include security of cyberphysical systems, power systems, model reduction, and fundamental limitations in control.
Dr. Sandberg received the Best Student Paper Award from the IEEE Conference on Decision and Control in 2004, an Ingvar Carlsson Award
from the Swedish Foundation for Strategic Research in 2007, and Consolidator Grant from the Swedish Research Council in 2016.
He has served on the editorial boards of IEEE TRANSACTIONS ON AUTOMATIC CONTROL and the IFAC Journal Automatica.
\end{IEEEbiography}




\end{document}